\documentclass[letterpaper]{article} 
\usepackage{aaai25}  
\usepackage{times}  
\usepackage{helvet}  
\usepackage{courier}  
\usepackage[hyphens]{url}  
\usepackage{graphicx} 
\usepackage{natbib}  
\usepackage{caption} 
\usepackage{algorithm}
\usepackage{algorithmic}
\usepackage{multirow}

\usepackage{xcolor}
\newcommand{\answerYes}[1]{\textcolor{purple}{#1}} 
 
\newcommand{\answerNA}[1]{\textcolor{gray}{#1}}

\usepackage{newfloat}
\usepackage{listings}
\DeclareCaptionStyle{ruled}{labelfont=normalfont,labelsep=colon,strut=off} 
\floatstyle{ruled}
\newfloat{listing}{tb}{lst}{}
\floatname{listing}{Listing}

\usepackage{booktabs}

\usepackage{amsmath}
\usepackage{tcolorbox}
\usepackage{enumitem}
\tcbuselibrary{listings}
\usepackage{tcolorbox}
\usepackage{afterpage}
\usepackage{cellspace}
\usepackage{subfigure}
\usepackage{pbox}

\newtcblisting{promptbox}[1]{%
  title=#1,
  listing only,
  colback=white!95!gray,
  colframe=gray,
  width=\columnwidth,
  arc=0mm,
  boxrule=0.5mm,
  colbacktitle=white!95!gray,
  coltitle=black,
  fonttitle=\bfseries,
  left=6pt,
  right=6pt,
  top=6pt,
  bottom=6pt,
  listing options={basicstyle=\small\ttfamily, breaklines=true}
}

\title{Extracting Participation in Collective Action from Social Media}
\author {
    Arianna Pera\textsuperscript{\rm 1},
    Luca Maria Aiello\textsuperscript{\rm 1,2}}

\affiliations {
    \textsuperscript{\rm 1}IT University of Copenhagen\\
    \textsuperscript{\rm 2}Pioneer Center for AI\\
    arpe@itu.dk, luai@itu.dk
}

\begin{document}

\maketitle

\begin{abstract}
Social media play a key role in mobilizing collective action, holding the potential for studying the pathways that lead individuals to actively engage in addressing global challenges. However, quantitative research in this area has been limited by the absence of granular and large-scale ground truth about the level of participation in collective action among individual social media users. To address this limitation, we present a novel suite of text classifiers designed to identify expressions of participation in collective action from social media posts, in a topic-agnostic fashion. Grounded in the theoretical framework of social movement mobilization, our classification captures participation and categorizes it into four levels: recognizing collective issues, engaging in calls-to-action, expressing intention of action, and reporting active involvement. We constructed a labeled training dataset of Reddit comments through crowdsourcing, which we used to train BERT classifiers and fine-tune Llama3 models. Our findings show that smaller language models can reliably detect expressions of participation (weighted F1=0.71), and rival larger models in capturing nuanced levels of participation. By applying our methodology to Reddit, we illustrate its effectiveness as a robust tool for characterizing online communities in innovative ways compared to topic modeling, stance detection, and keyword-based methods. Our framework contributes to Computational Social Science research by providing a new source of reliable annotations useful for investigating the social dynamics of collective action.
\end{abstract}

\section{Introduction}
The societal significance of online-mediated behavior change is growing, with grassroots, large-scale transformations being increasingly recognized as critical to addressing contemporary global challenges~\cite{van2009averting,otto2020social}.
While some online movements fail to yield tangible outcomes, in multiple occasions the Social Web has demonstrated its transformative potential in mobilizing collective action~\cite{gerbaudo2012tweets} to achieve various political, societal, and economic goals~\cite{yasseri2016political, boulianne2020young, lucchini2022reddit}.

Social movement theories suggest the presence of incremental pathways to collective action, where organized groups gradually mobilize individuals into becoming active \emph{participants}~\cite{van2013social}.
Analyzing these trajectories—a critical focus for many Computational Social Science studies on opinion dynamics and behavior change—remains challenging due to the scarcity of reliable, large-scale ground truth data.
The operationalization of social science theories relevant to collective participation, often overlooked in current literature on online behavior change~\cite{pera2023measuring}, could drive progress in this respect and support real-world applications.

To address these challenges, we combined insights from literature on mobilization and activism with modern Natural Language Processing (NLP) techniques to develop open-source text classifiers to detect expressions of participation in collective action from social media posts\footnote{Data and code are available at \url{https://github.com/ariannap13/extract_collective_action}.
Models are shared at \url{https://huggingface.co/ariannap22}.}.
Our classifiers identify the presence of participation and categorize it into one of four levels: (i) identifying collective problems and proposing solutions, (ii) engaging in calls-to-action, (iii) expressing intent to act, and (iv) reporting actual involvement in action.
The tools are \emph{topic-agnostic}, and therefore can be applied to large-scale platforms to generate extensive and granular ground truth data for various applications.

We evaluated four classification pipelines: a BERT-based classifier, a zero-shot Llama3 model, a Supervised Fine-Tuning (SFT) Llama3 model, and a Direct Preference Optimization (DPO) Llama3 model.
We relied on Reddit posts manually annotated via crowdsourcing for supervised learning. While expressions of participation can be relatively rare in social media—especially for the highest participation levels—we found that modern data augmentation techniques can aid the creation of high-quality data sufficient for training.

Overall, we answer the following research questions:

\vspace{2pt} \noindent \textbf{RQ1.} \emph{Can modern NLP reliably detect participation in collective action across topics?}

\vspace{2pt} \noindent \textbf{RQ2.} 
\emph{How can modern NLP models enhance existing methods and proxies to advance collective action analysis?}

\vspace{2pt} \noindent \textbf{RQ3.} \emph{Which Reddit communities exhibit higher levels of participation in collective action?}
\vspace{2pt}

Our contributions include: (i) a framework to quantify participation levels in collective action, adaptable across topics and contexts; (ii) the operationalization of such framework with a human-annotated dataset and advanced NLP tools; (iii) validation experiments and a case study showcasing our method's value compared to topic modeling, stance detection, and keyword-based methods tailored to capture topic-specific expressions of activism in language; and (iv) insights into how Reddit community demographics relate to collective action participation.   

We found that small models can reliably quantify participation in collective action, often rivaling Large Language Models (LLMs).
When applied to the binary classification task of identifying any expression of participation, the best BERT-based model achieved a macro F1 score of $0.65$, close to the $0.71$ of the best-performing LLM approach. In the multi-class classification of participation levels, the best BERT model obtained a macro F1 score of $0.44$, not far from the $0.52$ obtained by the best fine-tuned LLM.
Notably, while fine-tuned LLMs excel at nuanced classifications, they demand significantly more computational resources, with inference times up to two orders of magnitude larger than what smaller transformer-based models require. 
The choice of model should therefore balance resource availability with task requirements. 

Our proposed method differs from traditional topic modeling and stance detection, offering an effective tool for analyzing diverse data sources.
Applied to climate-related discussions on Reddit, it detects abundant expressions of participation in collective action in communities that would not be recognized as action-oriented by widely-used keyword-based proxies.
When extended to the broader Reddit platform, the method reveals higher levels of collective action participation in communities with audiences that are predominantly adult, female, and left-leaning.
Such insights demonstrate the potential of our approach to considerably broaden the scope of studies of collective action in online settings.

\section{Related Work}

\paragraph{Social Movements and Collective Action}
The study of social movements and collective action has long been of interest to social scientists, particularly in identifying participation predictors such as \emph{injustice}, \emph{efficacy}, and \emph{identity}~\cite{van2008toward, furlong2021social, thomas2022mobilise}. In online contexts, these theories have been operationalized using dictionary-based approaches~\cite{gulliver2021assessing, brown2022opposing} and network-based enhancements~\cite{erseghe2023projection}. In a similar context,~\citet{smith2018after} developed a dictionary of collective action terms which can be used to compute the percentage of words in text samples that align with these terms~\cite{pera2024narratives}.
However, these methods struggle to capture nuanced expressions of collective action and are limited by their reliance on fixed vocabularies, partial handling of word variations, and biases introduced by document length. 

\paragraph{Language Models in Text Classification}
Classifiers based on text embeddings have been used to extract social dimensions such as emotions, empathy~\cite{choi2020ten, choi2023llms}, and specific stages of collective action like calls-to-action~\cite{rogers2019calls}. While these methods provide greater flexibility, their performance is often constrained by their generalizability across contexts and scopes.

The advent of Large Language Models (LLMs) has created new possibilities for computational social science, especially in data-scarce scenarios~\cite{moller2024prompt}. Prior work has explored their utility in tasks such as stance detection~\cite{lan2024stance}, hate speech identification~\cite{kumarage2024harnessing}, and misinformation detection~\cite{huang2025unmasking}, often stressing the importance of hybrid approaches in such tasks.
However, less attention has been paid to the ability of LLMs to classify user participation in collective action, a task requiring a nuanced understanding of social movement dynamics. Unlike previous studies that focus on a single classification paradigm, our work systematically compares dictionary-based, centroid-based, BERT-based, and LLM-based classifiers to assess their effectiveness in capturing participation signals. This comparative approach allows us to evaluate generalizability across different modeling strategies, shedding light on the trade-offs between interpretability, accuracy, and computational cost in collective action classification. Specifically, while LLMs offer broad utility, their effectiveness relative to fine-tuned classifiers depends on the task and domain, with trade-offs in annotation and computational costs. In some cases, simpler fine-tuned classifiers outperform LLMs in specific tasks~\cite{ziems2024can}, underscoring the importance of task-specific model selection.

\paragraph{Studying Activism on Reddit}
Reddit has been a valuable source for analyzing group-specific characteristics, such as mental health~\cite{kim2023understanding} and political leanings~\cite{hofmann2022reddit}. However, individual engagement in activism through textual traces remains underexplored. Existing studies often use proxies like subreddit involvement to measure activism~\cite{lenti2025causal}, overlooking comment-level commitments to collective action.  In this work, we address this limitation by focusing on the extraction of comment-level participation in collective action, offering a new perspective on how online communities express engagement in activism.

\section{Theoretical Framework}

Our aim is to identify expressions of participation in collective action from natural language. To achieve this, we first need to define collective action and then identify the forms of participation that are relevant to the context of our study.

\subsection{Collective Action}
A well-established definition, rooted in psychology, describes collective action as any effort aimed at improving the conditions of an in-group, regardless of the number of participants or the outcome of the action~\cite{wright1990responding}.
However, such a definition has some limitations. It overlooks the critical link between collective action and group presence, focuses solely on collective motives while excluding normative and self-interest as plausible ones, and does not consider actions by non-members designed to improve the status or treatment of disadvantaged out-groups~\cite{wright2009next}. 
These limitations make the definition unsuitable for the context of online discussions, where individual choices and motives often intersect with collective processes, and users frequently advocate for the rights of others outside their immediate in-group.

Given the limitations of the mainstream definition and its restricted applicability to online spaces, we propose an alternative definition of collective action that integrates individual perspectives and acknowledges the role of the out-group.
We define collective action by first describing the nature of the collective action problem at its core. Specifically, we characterize a collective action problem as one that meets the following literature-informed criteria:
\begin{itemize}[leftmargin=*]
    \item \textbf{Human-generated}. The issue arises from social, political, or economic factors and affects a group of people (or an individual as a representative symptom of a broader issue) who are required to make choices in an interdependent situation~\cite{ostrom2010analyzing}.
    \item \textbf{Current}. The problem is either ongoing or rooted in the past but with clear and significant impact on the present or future. Problems confined to the past, with no ongoing or future implications, simply do not necessitate current action for their resolution.
    \item \textbf{With opportunity for action}. The issue can potentially be mitigated through collective efforts (e.g., rallies) or coordinated individual actions (e.g., recycling)~\cite{van2013four}.
    \item \textbf{Involving shared responsibilities}. Individuals, whether part of the in-group or out-group, can take actions that contribute to addressing the issue~\cite{van2009introduction}.
\end{itemize}
Based on this framework, we define \textbf{collective action} as any online or offline effort that can be taken to mitigate a collective action problem. 
Such mitigating actions can range from education and consciousness-raising, lobbying, negotiation, and petition signing to more disruptive activities such as boycotts, strikes and riots~\cite{wright2009next}.

\subsection{Mobilization to Define Participation in Action}
Mobilization is the process of allocating resources (such as time, money, or skills) to support a cause.
It serves as a prerequisite for collective action.
Existing theories on mobilization help categorize collective action stages, relating both to the individual agency to act and to the socio-psychological factors driving participation~\cite{klandermans1984mobilization}.
The literature on mobilization within social movements identifies the fundamental divide between non-action and action as a key conceptual distinction~\cite{wright2009next, klandermans2014people}. Operationalizing this distinction is especially valuable for applications aimed at quantifying collective action online, without focusing on differentiating action participation stages.

Understanding the transition from non-action to action requires examining two key processes: \emph{consensus mobilization} and \emph{action mobilization}.
\textbf{Consensus mobilization} refers to the organic convergence of views and meanings within social groups, driven by \emph{framing}—namely a set of active processes that construct a shared reality~\cite{klandermans1988formation}. There are three types of framing relevant to consensus mobilization~\cite{benford2000framing}: (i) diagnostic framing, defining problems; (ii) prognostic framing, proposing solutions; and (iii) motivational framing, creating a rationale for action.
\textbf{Action mobilization}, by contrast, focuses on promoting active participation~\cite{gamson1975strategy}. It unfolds through four steps~\cite{klandermans1987potentials}: (i) ensuring individuals sympathize with the cause, (ii) informing them about upcoming events, (iii) encouraging a willingness to act and (iv) removing barriers to ensure involvement.
Together, these two mobilization processes form a pathway for collective action, encompassing steps such as issue identification, solution proposals, calls to action, intention to act, and empowerment for active involvement. 

Traditionally, mobilization mechanisms have been studied through the lens of social movement organizations as mobilizing agents~\cite{snow1986frame}. In this work, we adopt the distinction between ``mobilization'' and ``participation'' proposed by~\citet{klandermans1987potentials} and shift the focus to the audience as the mobilized subject. This perspective applies the nuances of consensus and action mobilization to individuals, interpreting their steps toward participation as the grassroots counterpart to social movements' mobilization efforts.

We therefore categorize expressions of participation in collective action into four classes:
\begin{enumerate}[leftmargin=*]
    \item \textbf{Problem-solution}: Identifying a collective action issue, assigning blame, or proposing solutions.
    \item \textbf{Call-to-action}: Urging others to join or support a cause.
    \item \textbf{Intention}: Expressing willingness or interest in taking action.
    \item \textbf{Execution}: Reporting involvement in collective action initiatives.
\end{enumerate}
\noindent These categories are well-suited for studying online discourse and generalize across various collective action initiatives by identifying \emph{levels} of participation rather than specific activities tied to a particular cause.

\section{Methodological Framework}

We extracted and labeled comments from Reddit communities focused on activism and rights to create a training set for supervised learning. Comments were annotated on a multi-level scale: \emph{Problem-solution, Call-to-action, Intention}, and \emph{Execution}, which were collectively assigned the binary label \emph{Participation in collective action}. Comments not fitting these categories were labeled as \emph{None}.
We then trained models in two stages: first, to perform binary classification of participation, and second, to classify the specific level of participation.

\subsection{Data Curation} 
\paragraph{Data Collection}
We began our analysis by considering the top 40,000 subreddits ranked by all-time subscriber count, from which we identified a subset of 73 subreddits that contained the terms ``activism," ``activist," or ``rights" in their titles or descriptions. The identification of these subreddits was performed automatically by searching for the presence of these seed terms, which were selected for their direct relevance to the topic of collective action.
Following a manual review, we selected 42 subreddits pertinent to our study's scope, containing a total of 120 million comments posted from the creation of these subreddits through December 2023 (see Section~\ref{app:subreddits} in Appendix~\ref{app:data_model} for the exclusion criteria and the list of selected subreddits).

To refine our analysis towards comments more likely related to collective action, we adopted a word-matching approach. Specifically, we used an existing LIWC-like corpus comprising 47 terms associated with collective action~\cite{smith2018after}. We retained only comments with a minimum of two term matches. We then split the comments into sentences and keep the sentence with the highest frequency of these terms, along with one preceding and one following sentence to preserve the discussion context. For each subreddit, we split this final dataset of comments into possible candidates for training and test, totaling 2.7 million and 150,000 comments respectively.

\paragraph{Crowdsourcing}

We used Amazon Mechanical Turk (MTurk) to annotate comments. Workers were presented with one comment at a time and instructed to assign one label from \emph{Problem-solution, Call-to-action, Intention, Execution}, or \emph{None}. Label definitions and illustrative examples were provided (see Section~\ref{app:crowdsource} in Appendix~\ref{app:data_model}). Workers were instructed to prioritize higher levels of participation when multiple labels applied (e.g., \emph{Execution} over \emph{Intention}).

We selected a random sample of 50 comments from each of the 42 subreddits, resulting in 2,100 samples. Workers were paid \$0.18 per annotation, corresponding to a hourly rate of \$7.20 considering 90 seconds per annotation.

To ensure annotations of high quality, we added 672 control samples (30\% of real samples). These included 8 manually selected comments and 20 synthetically generated comments per participation label (i.e., excluding \emph{None}) using Llama3 8B Instruct~\cite{dubey2024LLaMa} (cf. Appendix~\ref{app:prompts} for the prompt). Control samples had clear, known labels to evaluate annotator performance.

Overall, we employed five quality-control strategies. 
First, we presented the tasks as images rather than HTML text to prevent easy task automation.
Second, we only recruited ``master" workers who had completed at least 5,000 annotations on MTurk with a minimum 95\% acceptance rate. 
Third, we discarded all annotations from workers failing more than 50\% of control samples.
Fourth, we removed samples with fewer than two annotators after filtering.
Finally, we excluded samples without a clear majority vote.
The final inter-annotator agreement measured with Krippendorff's alpha is 0.86.
The resulting annotated dataset consists of 369 samples: 100 labeled as \emph{None} and 269 as expressing \emph{Participation in collective action}, with 202 labeled as \emph{Problem-solution}, 44 as \emph{Call-to-action}, 14 as \emph{Intention}, and 9 as \emph{Execution}. Later on, we will refer to this dataset as \texttt{CS}.

\paragraph{Data Augmentation}

To address data sparsity in the annotated data, particularly in classes representing higher levels of participation, we applied two data augmentation techniques.

The first approach involved generating synthetic samples using Llama3 8B Instruct~\cite{moller24parrot}. Excluding the majority class (\emph{Problem-solution}), we generated 20 synthetic examples for each comment in the \emph{Call-to-action}, \emph{Intention}, \emph{Execution}, and \emph{None} classes. This dataset is referred to as \texttt{Syn A}. We also created a variation focusing solely on the minority classes \emph{Intention} and \emph{Execution} (\texttt{Syn I/E}). The generation prompt is provided in Appendix~\ref{app:prompts}. All generated samples were manually reviewed for accuracy using a pipeline based on the validity criteria from the prompt for each anchor example (e.g., semantic similarity, structure, meaning, intent, and key details). A table was created to binary-check the samples against these criteria.

The second method used Reddit's structural features. Assuming similar comments in analogous contexts share the same label, we excluded comments with a similarity higher than 0.95--likely content generated by bots--and retrieved the top 5\% most similar comments from the original discussion thread for each labeled comment. These were annotated with the same label as the original. Similarity was measured using Sentence-BERT~\cite{reimers2019sentence} with a rolling window to handle truncation issues. The concatenation of the crowdsourced annotated dataset and this Reddit-informed extension is referred to as \texttt{Ext}.

Table~\ref{tab:augmentation_volume} summarizes the label volumes for the crowdsourced dataset (\texttt{CS}), the synthetic augmentation for minority classes (\texttt{CS + Syn I/E}), the synthetic augmentation for all classes (\texttt{CS + Syn A}), the Reddit-informed extension (\texttt{Ext}), and the combination of both Reddit extension and synthetic augmentation for minority classes (\texttt{Ext + Syn I/E}).
\begin{table}[t!]
\centering
\begin{tabular}{@{}lccccc@{}}
\toprule
\textbf{Type} & \textbf{P-Sol} & \textbf{C-Act} & \textbf{Intent} & \textbf{Exec} & \textbf{None} \\ \midrule
\textbf{CS}              & 202 & 44  & 14  & 9   & 100 \\ 
\textbf{CS + Syn I/E}    & 202 & 44 & 275 & 182 & 100 \\ 
\textbf{CS + Syn A}    & 202 & 234 & 275  & 182  & 240 \\ 
\textbf{Ext}             & 553 & 175 & 24 & 16 & 285 \\ 
\textbf{Ext + Syn I/E}   & 553 & 175 & 285 & 189 & 285 \\ \bottomrule
\end{tabular}
\caption{N. of participation labels by augmentation type.}
\label{tab:augmentation_volume}
\end{table}

\paragraph{Test Set}
To construct the test set, we randomly sampled 20 comments from each of the 42 selected subreddits, using the pool of 150,000 comments designated for testing. To ensure diversity, we limited the sampling to one comment per unique author-discussion thread combination. Consequently, some subreddits may have fewer than 20 comments in the test set. Consistent with the training set methodology, we extracted the sentence with the highest frequency of action-oriented terms~\cite{smith2018after}, along with the preceding and following sentences. The final test set consists of 809 comments.

To ensure a high standard of annotation quality, we developed an annotation codebook (see Appendix~\ref{app:data_model}, Section~\ref{app:codebook}). This codebook further details the labeling guidelines established for the training set but it was deemed too complex for crowdworkers to use effectively within their time constraints. 
Both authors initially annotated 80 comments, achieving a Krippendorff's alpha of 0.69. Discrepancies, mainly between \emph{Problem-solution} and \emph{None}, were resolved through discussion. The first author then annotated the remaining samples.
In the end, 600 samples were labeled as \emph{None} and 209 as expressing \emph{Participation in collective action}, distributed as follows: 146 \emph{Problem-solution}, 39 \emph{Call-to-action}, 11 \emph{Intention}, and 13 \emph{Execution}.

\subsection{Classification Models}

We adopted a layered learning framework with a two-stage classification pipeline.
First, a \emph{binary classification} model determines whether a comment expresses participation in collective action. If participation is detected, a \emph{multi-class classification} model identifies its level. This framework is designed to optimize the accuracy of the initial binary classification stage, which is critical for downstream tasks.
This approach is common in multi-class classification with sparse labels~\cite{cerqueira2023automated} and theoretically aligns with the collective action literature, emphasizing the distinction between participation and non-participation~\cite{klandermans2014people} and accommodating potential use cases that may only necessitate a binary classification.

We evaluated four state-of-the-art Natural Language Processing (NLP) pipelines~\cite{moller2024prompt}: (i) a \emph{BERT-based} classifier fine-tuned on Reddit, (ii) a zero-shot Large Language Model (LLM) classifier, (iii) a Supervised Fine-Tuned LLM (SFT), and (iv) an LLM using Direct Preference Optimization (DPO). We also included word matching and centroid classification as
baseline models. Modeling details on training and prompt definitions are detailed in Section~\ref{app:model_details} of Appendix~\ref{app:data_model} and in Appendix~\ref{app:prompts}.

\paragraph{Dictionary (Dict)}
For binary classification, we used the dictionary of collective action words~\cite{smith2018after} to calculate the fraction of such words in the text. We then selected the threshold maximizing the difference between the True Positive Rate (TPR) and False Positive Rate (FPR) on the training set, prioritizing a class-balanced decision boundary to avoid under-detecting instances of ``Participation''. While we initially considered maximizing the F1-score, we found that it tended to favor the majority class, leading to a threshold that disproportionately benefited the more frequent ``None'' class. In contrast, maximizing $TPR - FPR$ allowed for a more balanced threshold, ensuring better detection of ``Participation'' cases without bias towards the majority class.

\paragraph{Centroid classifier (Centr)}
For both the binary and the multi-class tasks, we considered a centroid-based method, which has been used for similar classification tasks due to its simplicity and fast inference~\cite{park2020methodology}. This method computes the semantic similarity between a sample and the embedding centroid of text samples of each specific class, and selects the class with the closest centroid. We implemented it using the cosine similarity between the Sentence BERT embeddings of text samples~\cite{reimers2019sentence}.

\paragraph{BERT Classifier (BERT)}
We fine-tuned a RoBERTa model~\cite{liu2019roberta} on the full set of 2.7M Reddit comments reserved for training, with the aim of adapting the language model to recognizing linguistic patterns in Reddit comments related to collective action.
The model is then trained for both binary and multi-class classification tasks by considering a weighted loss function due to class imbalance.

\paragraph{Zero-shot LLM (ZS)}
In a zero-shot setting, a LLM classifies text based on its internal understanding of the task.
We designed the prompt using AI-enhanced knowledge prompting~\cite{liu2021generated}, leveraging Llama3 8B Instruct for label definitions and GPT-4 for prompt refinement. 
The classification of participation in collective action inherently relies on the definition of collective action and, by extension, the definition of a collective action problem. To explore this dependency, we experimented with prompts that either excluded or included a definition of the collective action problem (referred to as \texttt{no def} and \texttt{def}, respectively). 

\paragraph{Fine-tuned Llama Models}
We applied Supervised Fine-Tuning (\textbf{SFT}) to Llama3 8B Instruct by providing the model with a series of prompts that include: (i) fixed classification instructions specific to the task, along with all the allowed labels, and (ii) a set of labeled texts, derived from the crowdworkers' annotated dataset. 
We followed current best practices by adopting Quantized Low-Rank Adaptation (QLoRA), an efficient fine-tuning technique~\cite{dettmers2024qlora}. 
In this approach, the main model is frozen and quantized into a 4-bit representation, while separate low-rank matrices of gradients are learned during the fine-tuning process. 
These matrices are then combined with the frozen model during inference, weighted by a factor $\alpha$.

As an additional optimization step, we employed Direct Preference Optimization (\textbf{DPO}), a technique that updates the model’s weights based on explicit user preferences for one training example over another~\cite{rafailov2024direct}. 
During DPO, the model receives a prompt along with pairs of responses ranked by preference. The model’s weights are updated using cross-entropy loss to maximize the probability of generating the preferred example.

As with the zero-shot model, we tested prompts with and without a collective action problem definition (\texttt{no def} and \texttt{def}).

\section{Results}

\paragraph{Binary Task}

Table~\ref{tab:performance_binary} reports the performance of the binary classification task--the first step in our layered approach--measured in terms of F1 scores. Details on precision and recall values can be found in Table ~\ref{tab:performance_binary_precision_recall} in Appendix~\ref{app:results}.
The best-performing models were the zero-shot Llama3 model (both with and without the collective action problem definition in the prompt) and the RoBERTa classifier trained on the synthetically augmented training set. Across methods, the synthetic augmentation improved the classification performance considerably, up to $80\%$ for the macro F1 for RoBERTa. Fine-tuned LLMs underperformed simpler approaches. 

The computational time required for training and inference in LLM fine-tuning was significantly higher compared to RoBERTa. Using a Tesla V100 GPU, the SFT training took 1.5 to 4 hours to complete depending on the selected augmented training set, the DPO training required 12.5 to 24 hours and the fine-tuned RoBERTa training only 4 minutes.
Using a GTX 1080ti GPU, the average inference time per entry was 0.67s for the zero-shot approach, 1.09s for the fine-tuned LLM models and 0.007s for the RoBERTa classifier.
Given the comparably good performance, the sensitivity of LLMs to prompt changes, which can lead to inconsistencies~\cite{liu2023trustworthy}, the higher computational costs during inference and the complexity of hosting the models locally, we decided to select the RoBERTa model as the starting point for the multi-class step of the layered approach.

\begin{table}[t!]
\centering
\setlength{\tabcolsep}{2pt} 
\begin{tabular}{@{}c|l|ccccccc@{}}
 \textbf{Method}
                          & \textbf{Train. set}         & \textbf{Part.} & \textbf{None} & \textbf{\shortstack{Macro \\ Avg.}}
                          & \textbf{\shortstack{Weigh. \\ Avg.}}\\
\midrule
Dict.  &            & 0.39            & 0.51          & 0.45  & 0.48  \\
Centr. &
                          & 0.49            & 0.63          & 0.56   & 0.60       \\
\midrule
\multirow{2}{*}{{BERT}}  
                          & CS                      & 0.44            & 0.28          & 0.36  & 0.32              \\
                          & CS + Syn A                  & \textbf{0.51}   & \textbf{0.78} & \textbf{0.65} & \textbf{0.71}     \\
\midrule
\multirow{2}{*}{{ZS}} 
                          & no def                 & \textbf{0.56}   & \textbf{0.84} & \textbf{0.70} & \textbf{0.76}       \\
                          & def                 & \textbf{0.59}   & \textbf{0.83} & \textbf{0.71} & \textbf{0.77}        \\
\midrule
\multirow{4}{*}{{SFT}} 
                          & CS (no def)           & 0.47            & 0.50          & 0.49 & 0.49                \\
                          & CS (def)            & 0.48            & 0.47          & 0.47 & 0.47       \\
                          & CS + Syn A (no def)           & 0.50            & 0.59          & 0.54 & 0.57    \\
                          & CS + Syn A (def)           & 0.51            & 0.59          & 0.55 & 0.57       \\
\midrule
DPO 
                          & CS + Syn A (def)           & 0.47            & 0.42          & 0.44 & 0.43      \\
\end{tabular}
\caption{F1 scores for the binary classification task, with the top three models in \textbf{bold}. \texttt{CS} denotes the crowdsourced dataset, \texttt{CS + Syn A} indicates synthetic augmentation for all classes, and \texttt{def} and \texttt{no def} refer to whether the characteristics of a \textit{collective action problem} are defined in the LLM prompt or not.}
\label{tab:performance_binary}
\end{table}
\begin{table*}[t!]
\centering
\begin{tabular}{@{}c|l|ccccccc@{}}
\textbf{Method}    &  \textbf{Train. set}    & \textbf{Problem-solution} & \textbf{Call-to-action} & \textbf{Intention} & \textbf{Execution} & \textbf{None} & \textbf{\shortstack{Macro \\ Avg.}} & \textbf{\shortstack{Weight. \\ Avg.}} \\
\midrule
Centr. &             & 0.38                      & 0.15                    & 0.04               & 0.11                   & 0.78          & 0.29  & 0.66                \\
\midrule
\multirow{2}{*}{{BERT}} 
                          & CS      & 0.36                      & 0.56                    & 0.11               & 0.25                   & 0.78          & 0.41 & 0.68               \\
                          & CS + Syn A      & 0.36                     & 0.62                   & 0.17               & 0.29                  & 0.78       & 0.44 & 0.68             \\
\midrule
\multirow{2}{*}{{ZS}} 
                          & no def                & 0.34                     & 0.54                    & 0.09               & 0.46                   & 0.79          & 0.44 & 0.68                \\
                          & def                 & 0.38             & 0.39                    & 0.09      & 0.36                   & 0.78          & 0.40 & 0.67                \\
\midrule
\multirow{5}{*}{{SFT (def)}} 
                          & CS              & 0.37                      & 0.56                    & 0.13               & 0.43                   & 0.78          & 0.45 & 0.68                \\
                          & Ext                   & 0.35                      & 0.38                    & 0               & 0                   & 0.78          & 0.3 & 0.66                \\
                          & Ext + Syn I/E & 0.36                      & 0.39                    & 0.2               & 0.3                  & 0.78          & 0.41 & 0.67                \\
                          & CS + Syn I/E & 0.36                      & 0.55                    & 0.17               & 0.42                   & 0.78          & 0.46 & 0.68                \\
                          & CS + Syn A & \textbf{0.37}                      & \textbf{0.64}                    & \textbf{0.29}               & \textbf{0.5}                   & \textbf{0.78}          & \textbf{0.52} & \textbf{0.69}                \\
\midrule
\multirow{5}{*}{{DPO (def)}} 
                          & CS              & 0.28                      & 0.17                    & 0               & 0.29                   & 0.78          & 0.30 & 0.64                \\
                          & Ext                  & 0.34                      & 0.44                    & 0               & 0                   & 0.78          & 0.31 & 0.66                \\
                          & Ext + Syn I/E     & 0.34             & 0.26           & 0               & 0.12          & 0.78 & 0.3 & 0.65      \\
                          & CS + Syn I/E & \textbf{0.37}                      & \textbf{0.62}                    & \textbf{0.21}              & \textbf{0.52}                   & \textbf{0.78}         & \textbf{0.5} & \textbf{0.69}               \\
                          & CS + Syn A & \textbf{0.38}                      & \textbf{0.58}                    & \textbf{0.29}               & \textbf{0.5}                   & \textbf{0.78}          & \textbf{0.51} & \textbf{0.69}                \\
\end{tabular}
\caption{F1 scores for the multi-class classification task as a second step of the layered approach, with the top three models in \textbf{bold}. \texttt{CS} denotes the crowdsourced dataset, \texttt{CS + Syn I/E} indicates synthetic augmentation for minority classes, \texttt{CS + Syn A} for all classes, \texttt{Ext} denotes the Reddit-informed extension, and \texttt{Ext + Syn I/E} defines the combination of both Reddit extension and synthetic augmentation for minority classes. All LLM prompts define a \textit{collective action problem}.}
\label{tab:performance_multiclass}
\end{table*}
\paragraph{Multi-class Task}

In the second step, we trained classification models to identify the four levels of participation in collective action. During testing, we focused on comments selected in the first classification stage by the BERT-based classifier trained on synthetically augmented data.

Table~\ref{tab:performance_multiclass} reports performances for the multi-class task in terms of F1 scores, considering the inclusion of the collective action problem definition in the LLM prompts. Details on precision and recall values can be found in Table ~\ref{tab:performance_multiclass_precision_recall} in Appendix~\ref{app:results}. The SFT Llama3 model, trained on crowdworker-annotated data with synthetic augmentation for all classes except the majority class (\texttt{CS+Syn A}), achieved the best overall performance. The DPO Llama3 model, also trained on annotated data with the two synthetic augmentation strategies, followed closely.
Given that prompt variations and the limited number of \emph{Intention} and \emph{Execution} labels in the test set affect reliability for these classes, the RoBERTa classifier, trained on synthetically augmented data, remains a robust choice for this stage. However, the SFT and DPO models consistently outperformed others, particularly in classifying the minority classes \emph{Intention} and \emph{Execution}.

The ranking of models remained largely consistent whether or not the collective action problem definition was included in the LLM prompts or when models were applied directly to the multi-class task without a layered approach (see Appendix~\ref{app:results}, Tables~\ref{tab:performance_multiclass_direct} and~\ref{tab:performance_multiclass_nodef}).

\section{Validation}

We conducted four experiments to validate our proposed method. In particular, we focused on the first step of the layered pipeline, which is crucial to the overall classification task, and consider the BERT-based classifier trained on synthetically augmented data as reference. First, we compared our method to topic modeling to show that expressions of participation in collective action go beyond mere thematic categorization. Second, we contrasted our approach with stance detection, a task focused on identifying a position on a specific issue. Our method differs in its applicability across issues and in the fact that it focuses on behaviors rather than on opinions.
Third, as our training sample selection was based on the collective action words dictionary~\cite{smith2018after}, we examined the impact of this choice on the final classification results.
Finally, we evaluated the binary classification of participation in collective action on a dataset that is distinct from Reddit.
\begin{figure*}[t!]
  \centering
  \subfigure[Datasets\label{fig:umap_datasets}]{\includegraphics[width=0.16\textwidth]{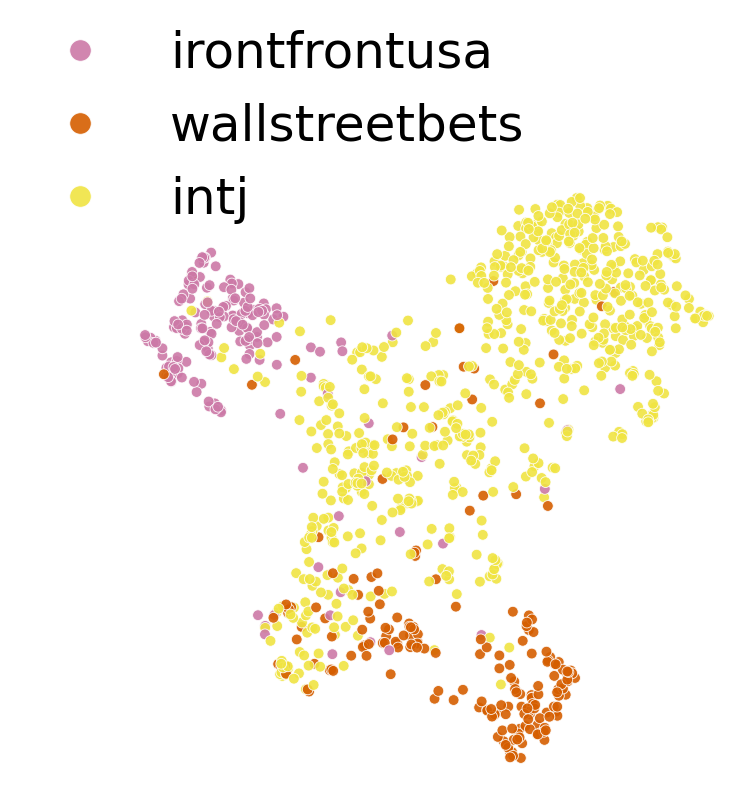}}
  \hfill
  \subfigure[Topics\label{fig:umap_topics}]{\includegraphics[width=0.15\textwidth]{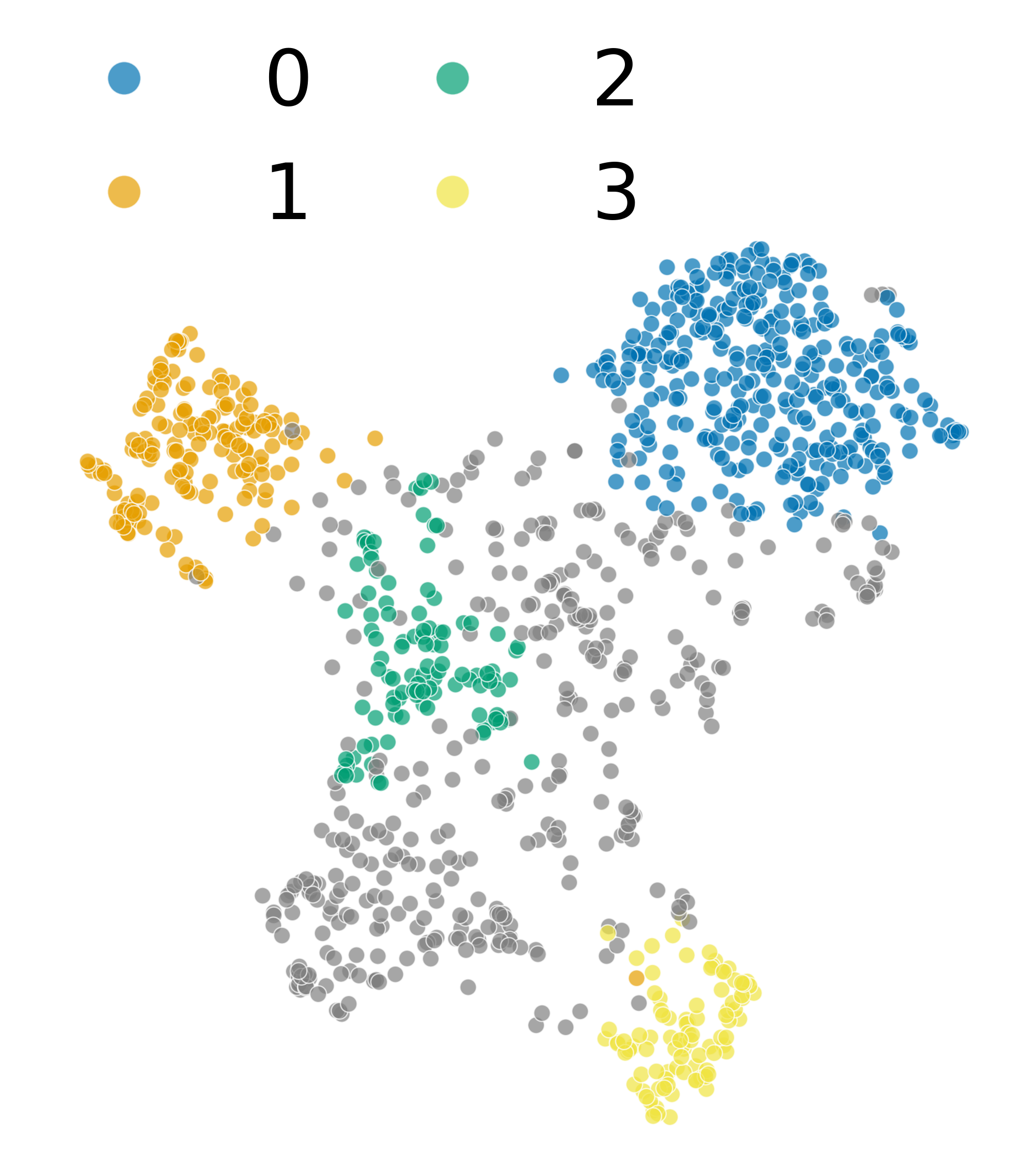}}
  \hfill
\subfigure[Commitment\label{fig:umap_commitment}]{\includegraphics[width=0.16\textwidth]{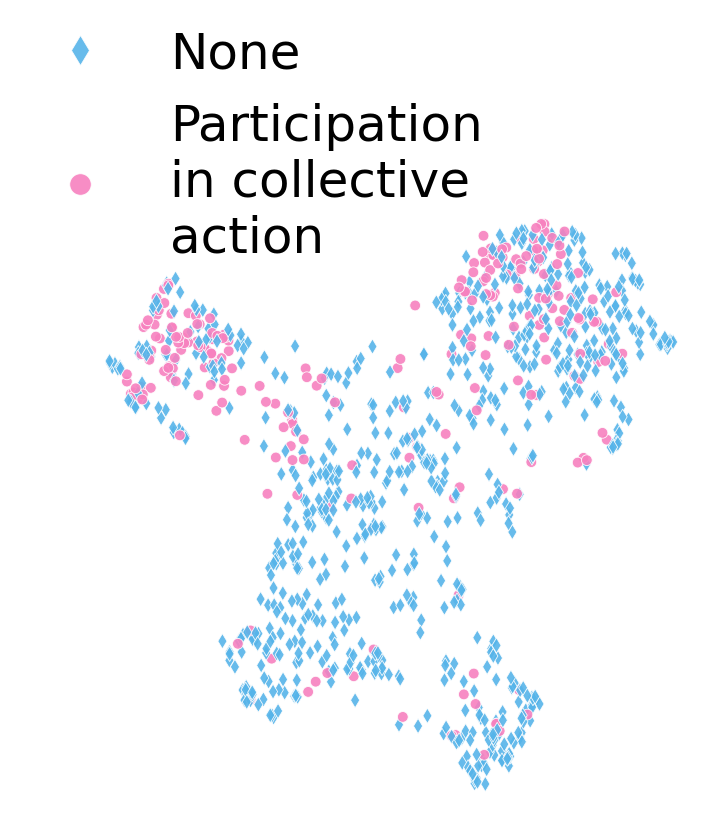}}
\hfill
  \subfigure[Test Set Topics\label{fig:perc_comm_topics_testset}]{\includegraphics[width=0.18\textwidth]{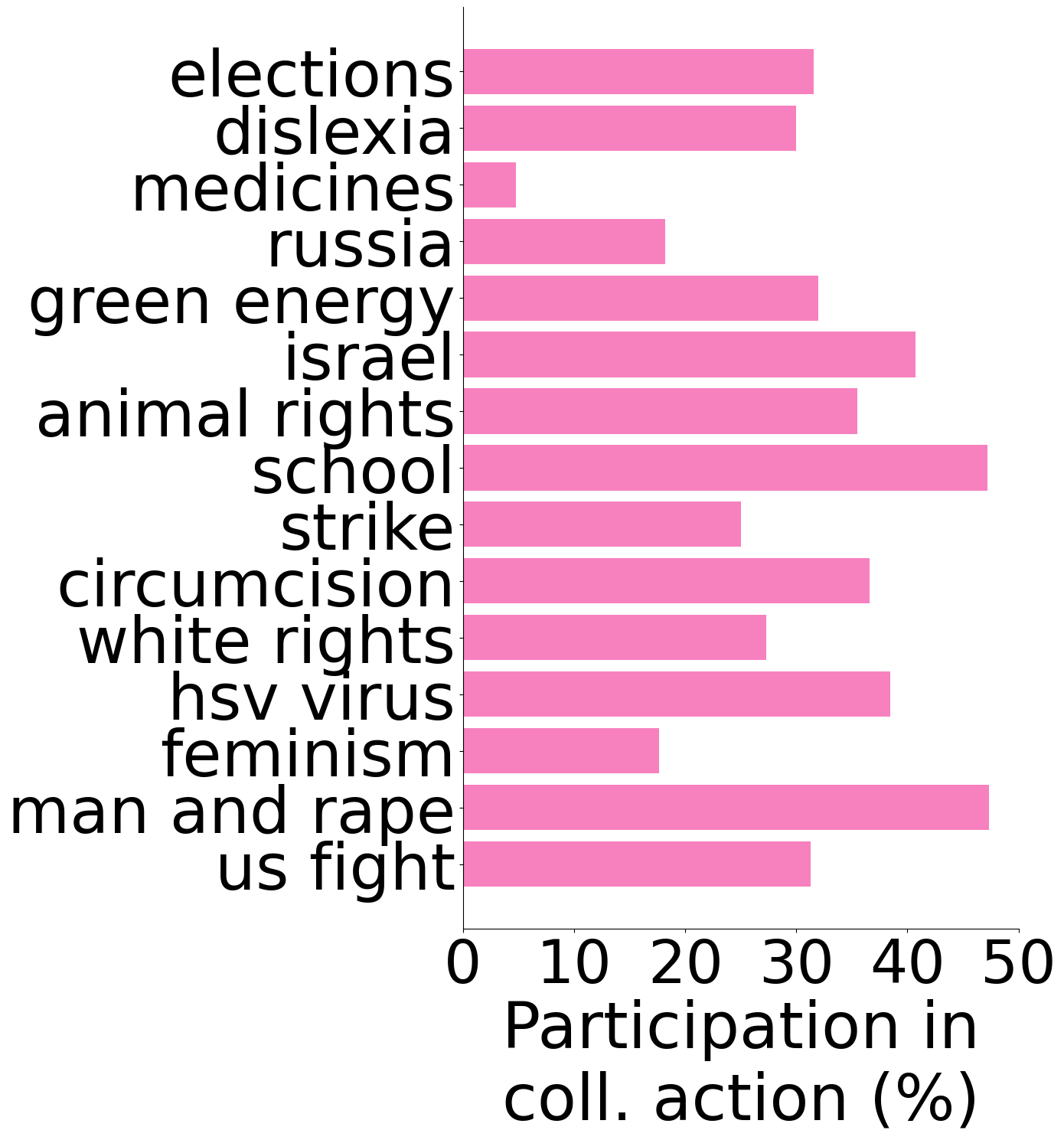}}
  \hfill
\subfigure[Stance\label{fig:stance_commitment}]{\includegraphics[width=0.15\textwidth]{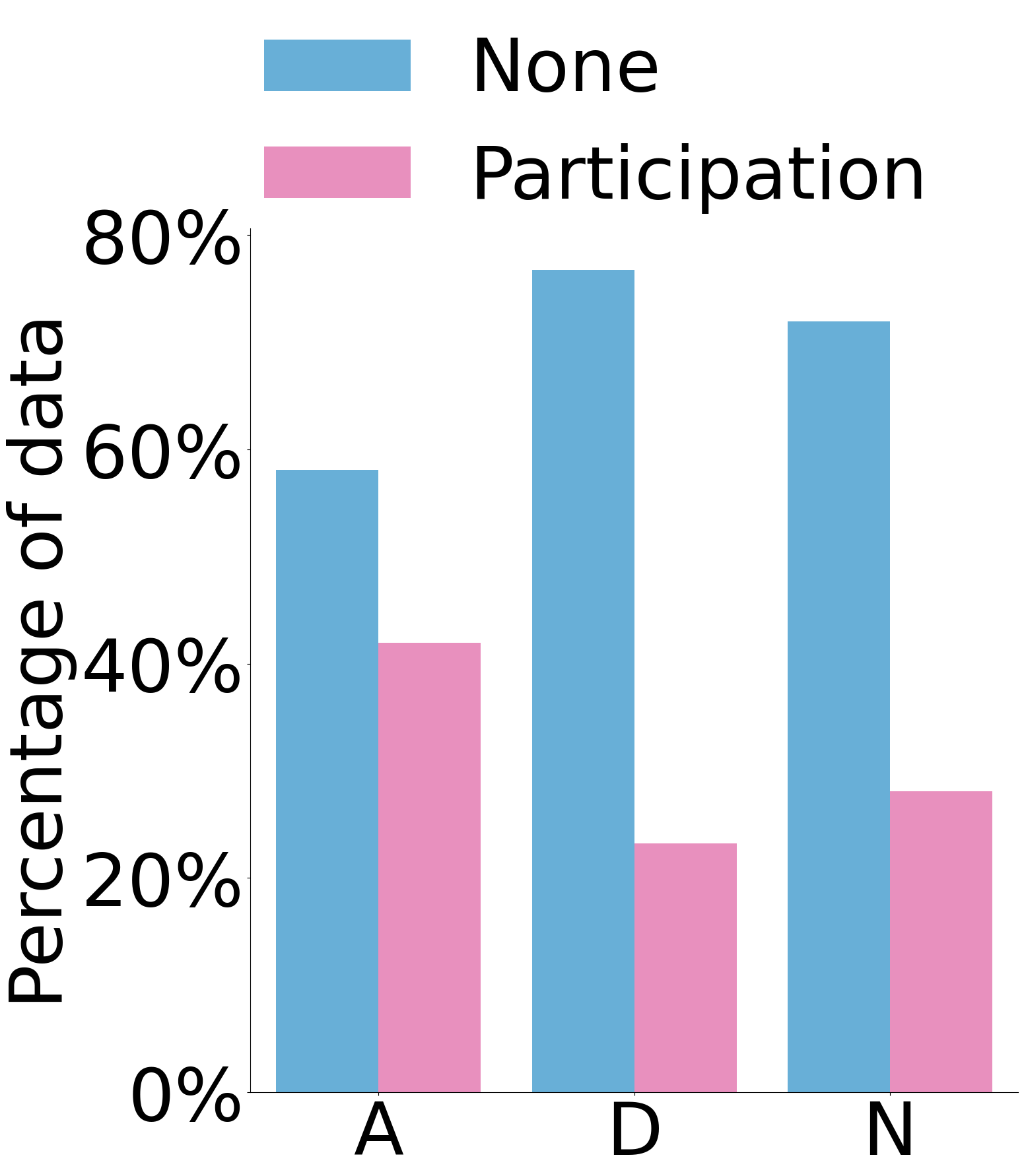}}
  \hfill \subfigure[Keywords\label{fig:keywords_influence}]{\includegraphics[width=0.15\textwidth]{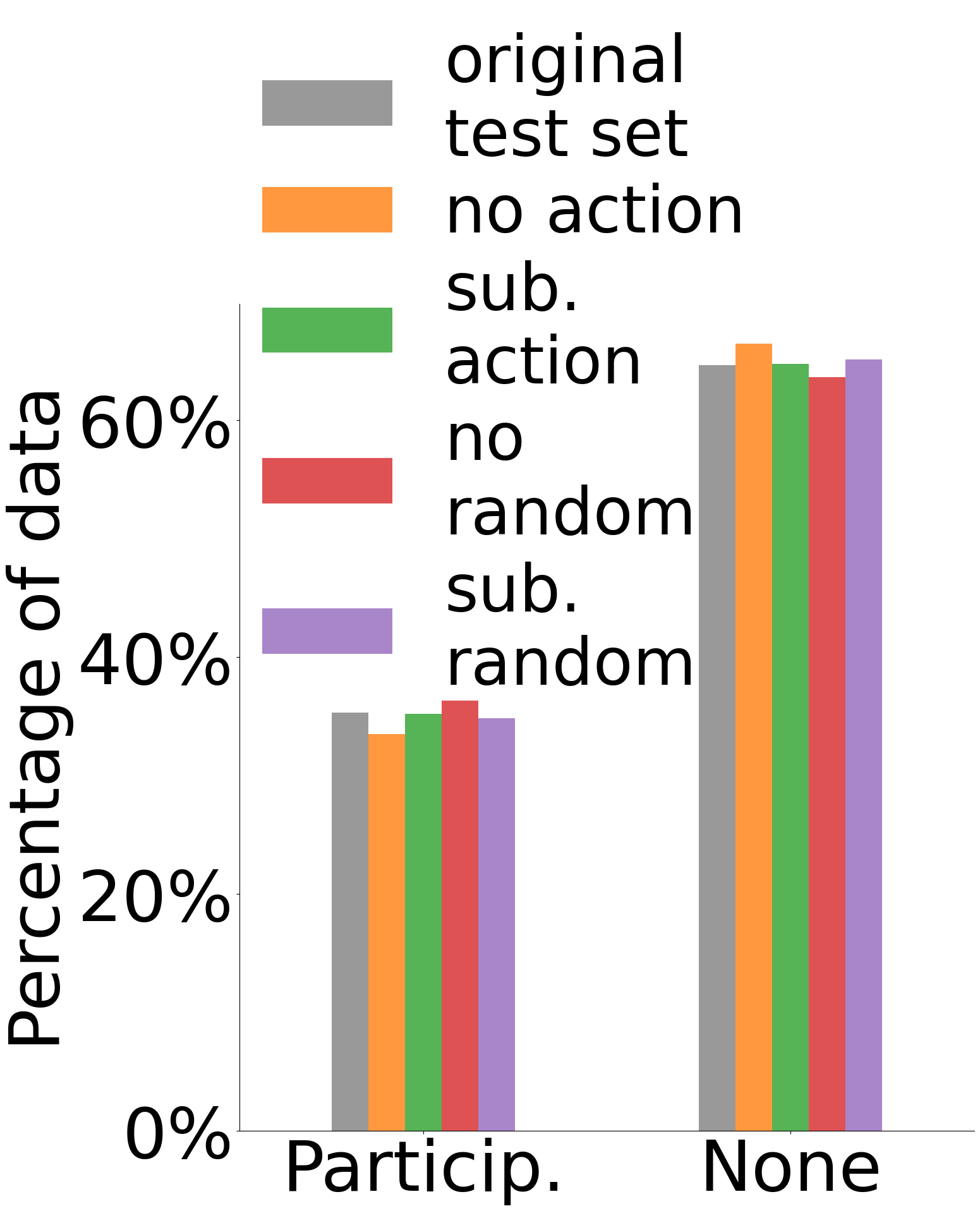}}
    \caption{Validation of the proposed approach. (a-d) comparison with topic modeling; (e) comparison with stance detection -- A for agree, D for disagree, N for neutral; (f) impact of collective action keywords on classification performance.}
\end{figure*}
\paragraph{Topic Modeling}
Our approach is (i) topic-agnostic, not distinguishing between different collective action topics, and (ii) capable of detecting participation in collective action even in subreddits not included in the training set. To demonstrate this, we analyzed three Reddit threads relevant to collective action: one from \emph{r/intj} on COVID-19 vaccine intake (``Did you get the COVID vaccine? Why or why not?", 679 comments), one from \emph{r/wallstreetbets} on class action lawsuits in the GameStop short squeeze (``Robinhood App Hit With Class Action Suit After Trying to Shut Down GameStop Uprising", 200 comments), and one from \emph{r/IronFrontUSA}--part of our training set--on the importance of voting (``Posting this again to remind people that if you refuse to vote [...] you share a huge amount of responsibility for what just happened", 180 comments).

We collected all comments from these threads and applied BERTopic~\cite{grootendorst2022bertopic} to a shuffled concatenation of them. We performed a minor text pre-processing by removing English stopwords and the 20 most frequent words in the whole collection. 
This resulted in nine distinct topics (plus noise), of which the largest ones are: \emph{covid} and its \emph{effects} (332 comments), \emph{voting} and \emph{democrats} (165 comments), \emph{science} and \emph{intjs} (141 comments), \emph{robinhood}, \emph{money}, and \emph{GameStop} (105 comments) (cf. Figure~\ref{fig:umap_topics}). The remaining five topics are significantly smaller in size (with the biggest containing 31 comments) (cf. Table~\ref{tab:topics_full} and Figure~\ref{fig:umap_topics_all} in Appendix~\ref{app:results}). 

If our method were not topic-agnostic, Figure~\ref{fig:umap_commitment} would show a distribution similar to Figure~\ref{fig:umap_datasets}. However, comments expressing participation in collective action do not follow a clear dataset dependency, although participation is featured less prominently in the \emph{r/wallstreetbets} thread.

If our method's detection capability were limited to its training set, no comments expressing participation would be detected in \emph{r/intj} or \emph{r/wallstreetbets}. However, at least some comments expressing participation were identified in all analyzed subreddits (see Figure~\ref{fig:umap_commitment}).

Additionally, if our method resembled topic modeling, we would expect topics extracted from the reference test set using BERTopic to be predominantly classified as either full participation in collective action or \emph{None}. However, among the 14 identified topics, most are nearly evenly split between participation and \emph{None} (see Figure~\ref{fig:perc_comm_topics_testset}).

\paragraph{Stance Detection}
In addition to being applicable across various topics, we demonstrate that our approach differs from stance detection, particularly in how stance classes relate to participation in collective action. To test this, we used a stance detection dataset related to global warming~\cite{luo2020detecting}. This dataset includes sentences from news articles annotated for their stance towards the statement ``Climate change/global warming is a serious concern," with labels \emph{agree}, \emph{disagree}, and \emph{neutral}.

We applied our binary classification approach to the dataset and analyzed the distribution of participation labels across stance classes. If our method were akin to stance detection, we would expect each stance class to align predominantly with either full participation or \emph{None}. However, as shown in Figure~\ref{fig:stance_commitment}, while there is some correlation between stance and participation, participation is detected in all stance classes at varying levels, indicating that stance detection cannot fully replace our method. 

\paragraph{Impact of Collective Action Terms}
We used the collective action words dictionary~\cite{smith2018after} to filter comments in both the training and test sets, as well as to extract relevant portions of each comment for classification models and annotators. To ensure our classifier does not rely solely on the presence of these words, we tested the classifier on four variations of the test set: (i) removal of action dictionary words, (ii) substitution of action dictionary words with random tokens from the model vocabulary, (iii) removal of random words, and (iv) substitution of random words with random tokens. This approach aligns with established practices in NLP for evaluating model robustness using adversarial datasets~\cite{ebrahimi2017hotflip, ghaddar2021context}.

If collective action words influenced the classifier, we would expect different results for configurations (i) and (ii) compared to the original test set. However, as shown in Figure~\ref{fig:keywords_influence}, the distribution of participation does not significantly differ under any of the tested conditions.

\paragraph{Classifying Parliamentary Debates}
To demonstrate the broader applicability of our framework, we applied the classifier to detect participation in collective action in a subset of UK parliamentary debate transcripts~\cite{ukdebates}. We selected debates from March 22nd, 25th, and 26th, 2024, totaling 1,000 text samples. Our approach effectively distinguishes messages expressing participation from those that do not, either by being unrelated to collective action or not identifying the issues as problematic (Table~\ref{tab:examples_debates}). While this analysis is preliminary, it demonstrates the potential of our approach for structured political discourse, highlighting a promising avenue for future research on collective action detection in institutional settings.
\begin{table}[t!]
\centering
\begin{tabular}{@{}c|p{0.9\columnwidth}@{}}
\toprule
\rotatebox[origin=c]{90}{} 
                          & \textbf{Text} \\
\midrule
\multirow{11}{*}{\rotatebox[origin=c]{90}{High probability}}
                          & 1. There is a desperate need for increased humanitarian support to Gaza. The UK, including the Ministry of Defence, is working collectively with allies [...]  \\
                          & 2. The Israeli Government have said that they want to “flood” Gaza with aid. Will my right hon. Friend assure the House that we will work with our partners globally to get more aid [...] \\
                          & 3. [...] the Prime Minister and I have been very proactive in speaking to and making multiple visits to the region. [...] There is now a large-scale programme of using a pier to get food in [...]\\
\midrule
\multirow{9}{*}{\rotatebox[origin=c]{90}{Low probability}}
                          & 1. I said that we are not going to be partisan in this debate, and the shadow Minister started in that vein, but my right hon. [...]  \\
                          & 2. The hon. Gentleman’s point is made more potent by the fact that the matters the ISC considers are not typically—in fact, not at all—partisan. [...] \\
                          & 3. How many prisoners have been released early under the end of custody supervised licence scheme since October 2023.\\
\bottomrule
\end{tabular}
\caption{Selected examples of parliamentary debates transcripts from the top and bottom 10 probabilities of class \emph{Participation in collective action}.}
\label{tab:examples_debates}
\end{table}
\begin{figure}[t!]
    \centering   
    \includegraphics[width=0.47\textwidth]{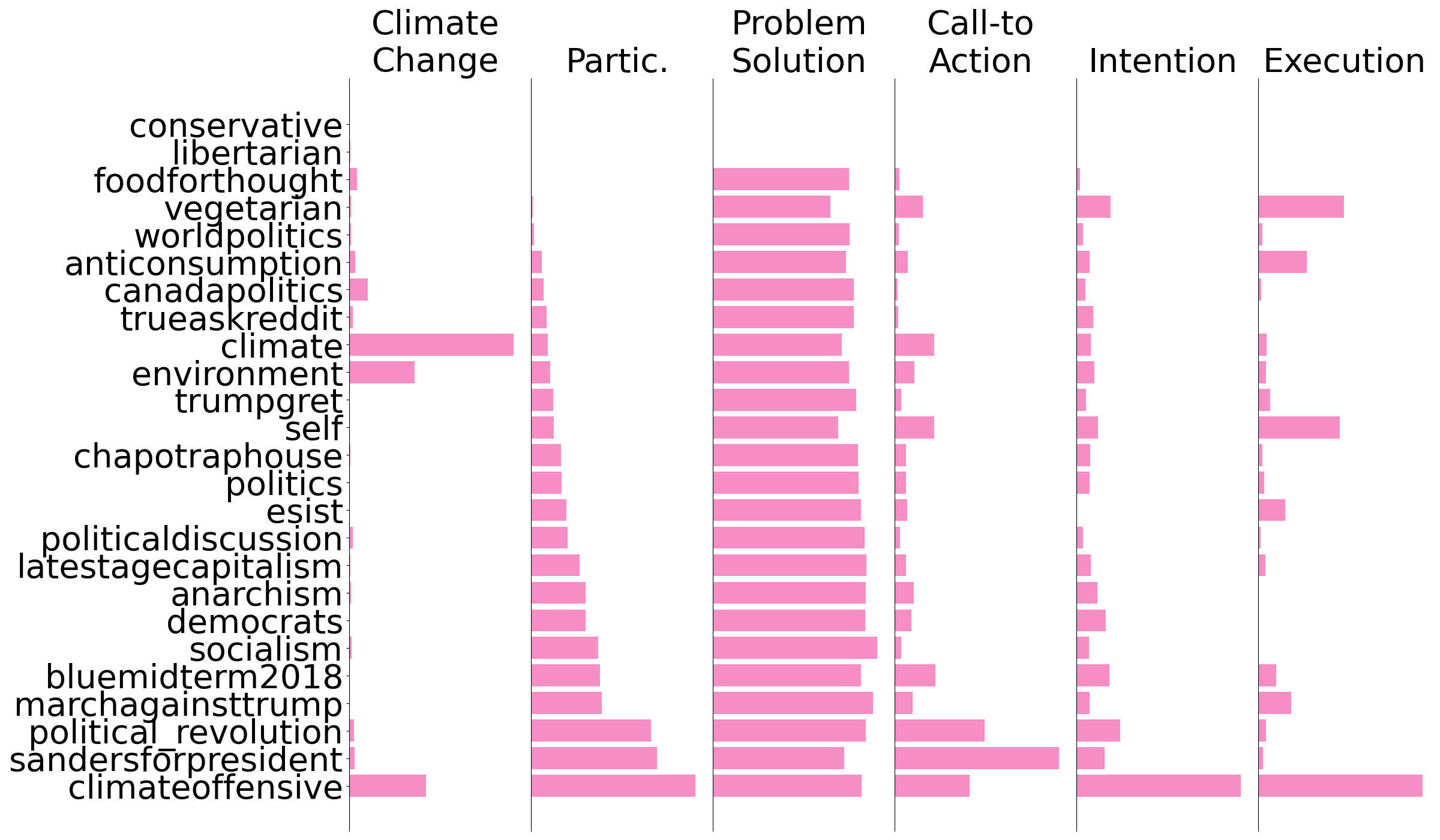}
    \caption{Comparison of climate change-focused comment percentages and participation in collective action levels across subreddits, x-axis values normalized.}
    \label{fig:climatechange_committment}
\end{figure}
\section{Climate Action}
We now demonstrate the effectiveness of our method for detecting comments indicative of participation in collective action, comparing it to widely used approaches in the literature. We used climate action as a case study due to the extensive research in this area, particularly within Reddit~\cite{oswald2022climate, treen2022discussion}. 
Previous studies have largely relied on subreddit selection and keyword filtering, assuming that the presence of certain keywords or subreddit participation directly reflects participation in climate action~\cite{parsa2022analyzing, lenti2025causal}. In contrast, we show that subreddits selected based on keyword counts may not always align with actual participation in collective action.

We began with a potential set of the top 2,000 all-time subreddits ranked by subscriber count. Using a list of climate change-related substrings from~\cite{parsa2022analyzing} and focusing on 2018, we filtered this set to include only subreddits containing climate-related comments, resulting in 1,597 subreddits. To further enhance our dataset, we added additional subreddits identified as relevant by~\citet{parsa2022analyzing} that were not part of the initial set, bringing the total to 1,609 subreddits.
From these communities, we collected 719,667 keyword-matching comments posted in 2018, which were then analyzed with our best BERT-based binary classifier to detect those expressing participation in collective action.
To streamline our analysis, we focused on 440 subreddits with at least 100 climate change-related comments during the specified timeframe. These subreddits were ranked based on two criteria: (i) the fraction of comments containing climate-related keywords and (ii) the fraction expressing a participation in collective action alongside climate change-related keywords. The rankings showed a moderate positive but weak correlation (Spearman $R=0.34$), with notable divergence at the top positions.
Figure~\ref{fig:climatechange_committment} highlights 24 subreddits where at least 40\% of comments express participation in action, ranked by ascending percentage of such a measure. Subreddits like \emph{r/climate} and \emph{r/environment}, while highly ranked by climate-related mentions, did not appear in the top 10 for participation in collective action. Conversely, subreddits such as \emph{r/marchagainsttrump}, \emph{r/bluemidterm2018}, and \emph{r/democrats} ranked very low for climate mentions but show a relatively high frequency of participation in collective action. Depending on the specific type of participation in collective action, detected through Llama3 fine-tuned on synthetically augmented data, different subreddits may be more or less suitable as data sources. For instance, \emph{r/ClimateOffensive, r/self, r/vegetarian} and \emph{r/anticonsumption} were the highest ranked in terms of \emph{Execution} among the 24 visualized communities.
This case study demonstrates how our framework identifies online communities relevant to climate activism more effectively than coarse metrics like subreddit popularity or keyword frequency.

\section{Sociodemographic Dimensions of Participation}
With the rise of online media in public discourse~\cite{boulianne2020young}, research increasingly seeks to characterize the socio-political attributes of social media users through embedding-based methods that distill semantic information from user content to infer attributes like age, income, and partisanship~\cite{aceves2024mobilizing}. These embeddings have been used to study political engagement by correlating key dimensions with outcomes such as collective decision-making, online mobilization, and advocacy for global causes~\cite{israeli2022must, kumar2018community, lenti2025causal}. However, such studies often rely on coarse proxies, such as subreddit participation as a stand-in for activism on Reddit~\cite{lenti2025causal}.

Our classification pipeline presents a new opportunity to examine whether the socio-political dimensions estimated by state-of-the-art embedding methods correlate with \emph{direct expressions} of participation in collective action, and whether these correlations align with theoretical expectations.
Specifically, we tested three hypothesis: (i) younger generations contribute significantly to activism, though their involvement may differ from older generations~\cite{earl2017youth}, (ii) women and men tend to engage differently in political activism~\cite{coffe2010same}, and (iii) awareness of wealth inequality can drive people toward activism, with liberals more likely to engage in activism aimed at reducing inequalities~\cite{hoyt2018wealth}.

To test these hypotheses, we used Reddit community embeddings~\cite{waller2019generalists} to characterize the socio-political position of subreddits.
For the top 1,000 subreddits by all-time subscriber count, we combined these embeddings with the social dimensions defined by~\citet{waller2021quantifying} to compute scores estimating communities' standing along three axes: age (young/old), gender (male/female), and partisanship (left-wing/right-wing). These scores were then compared to the fraction of comments expressing explicit participation in collective action, predicted through our BERT-based classifier trained on synthetically augmented data. The embeddings were derived from comments posted between 2005 and 2018 and, to meet computational constraints, the collective action analysis used a random 5\% sample of comments published in May 2018, totaling 1.4 million.

\begin{figure}[t!]
    \centering
    \includegraphics[width=0.47\textwidth]{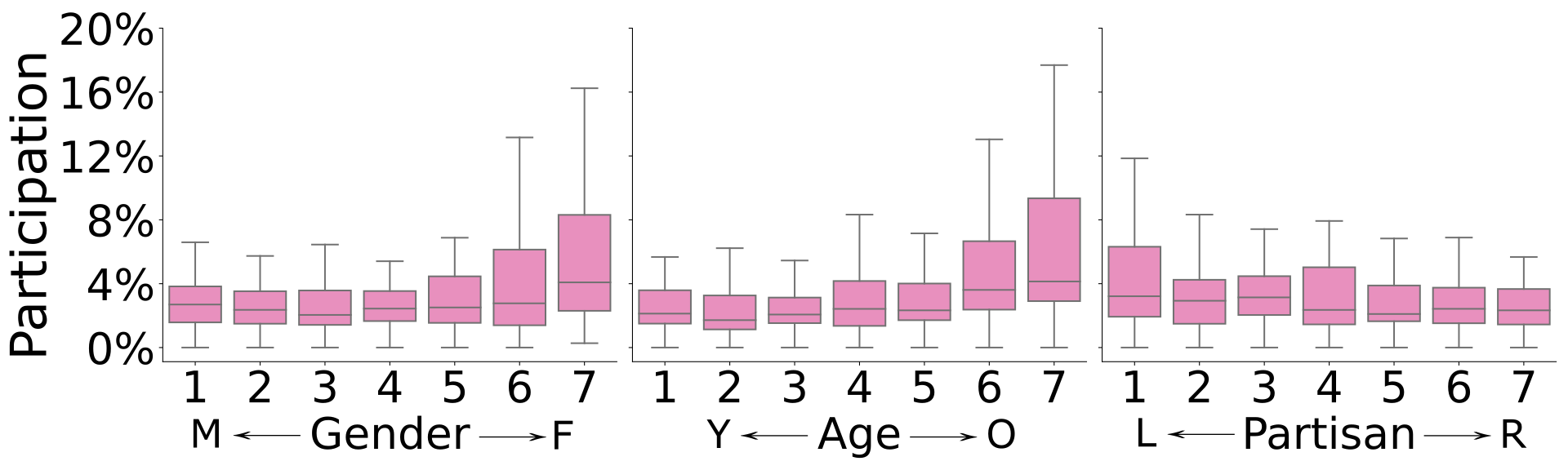}
    \caption{Fraction of Reddit comments showing participation in collective action across subreddits with varying socio-political tendencies. The socio-political scores are divided into equally sized bins based on sample quantiles.}
    \label{fig:commitment_social_binary}
\end{figure}

While variability across subreddits is substantial, Figure~\ref{fig:commitment_social_binary} reveals three notable patterns.
First, subreddits aligned more with the female pole show a higher fraction of comments expressing participation in action, supporting the hypothesis that gender is a dimension that partly explains activism patterns.
Second, subreddits skewed towards older demographics also exhibit a higher fraction of such comments, suggesting meaningful contributions from older generations.
Note that ``older'' here reflects platform-relative demographics and does not contradict the hypothesis that younger generations are highly active, considering that 64\% of Reddit users are under 30 years old~\cite{barthel2016reddit}.
Last, there is a weaker, inverse relationship between conservative-leaning subreddits and the prevalence of participation-related comments, consistent with the hypothesis that liberals are more likely to engage in activism.

\section{Discussion and Conclusion}
Drawing inspiration from the literature on collective action and mobilization, we contribute to the Computational Social Science community with new open-source text classifiers designed to accurately extract expressions of participation in collective action across topics.
Our classification pipeline can operate in a layered fashion to capture different levels of participation.
We achieve this by employing state-of-the-art Natural Language Processing models through a supervised learning approach, complemented by modern data augmentation techniques.
We show that relatively high classification accuracy is attainable even with a limited number of manually-labeled examples.
In our analysis of 61 different combinations of models and data augmentation strategies, we discovered that transformer-based small language models can compete effectively with more modern Large Language Models.
Notably, our BERT-based classifier offers an optimal balance between performance, computational efficiency, and cost-effectiveness across both layers of classification (\textbf{RQ1}).
While the specific reasons behind the comparable performance of small language models and fine-tuned LLMs on our binary classification task remain unclear, similar findings have been reported in both NLP and Computational Social Science literature~\cite{bucher2024fine, ziems2024can}. This suggests that smaller, domain-specific models can rival larger models for certain tasks, and we consider this an important avenue for future research.

In applying our tool to data derived from Reddit discussions, we observed that our classification yields results that are orthogonal to those produced by traditional topic modeling and stance detection algorithms.
When applied to climate-related communities, our tool reveals a profile of participation in collective action that differs significantly from the results obtained through conventional keyword matching methods.
This highlights the potential of our approach to improve the text analytics pipeline over topic-specific approaches and coarser proxies of collective action previously employed in the literature (\textbf{RQ2}).
Furthermore, our analysis indicates that discussions rich in expressions of participation in collective action are prevalent in forums characterized by more female, adult and left-leaning participants, in line with theoretical expectations (\textbf{RQ3}).

\paragraph{Limitations and Future Work}
Our proposed method has some limitations.
First, our focus on a limited number of seed terms—``activism," ``activist," and ``rights"—may have biased the selection of subreddits, potentially overlooking relevant but ideologically diverse communities. This could skew the dataset toward left-leaning discourse, affecting downstream applications. Future work could refine subreddit selection by incorporating additional terms or applying more systematic, politically balanced criteria.
Second, the volume of the annotated training set is relatively small. 
While we have validated the augmented data and used it strategically to mitigate the cost of crowdsourcing annotations, its influence on model performance and potential biases warrants further investigation. This is particularly important given class imbalance and the presence of rare classes, underscoring the need for future research on their treatment and the development of tailored augmentation strategies to ensure robustness and reliability.
Third, while the training set consists of crowdsourced annotations, our test set is annotated by domain experts.
Although we prioritized the accuracy and robustness of the test set, incorporating expert-annotated data into the training set could bolster its reliability. 
Fourth, while we framed the classification of nuanced levels of participation in collective action as a multi-class task, it could alternatively have been approached as a multi-label task, given that multiple levels may apply to a single text. Fifth, our experiments are primarily based on Reddit data, with the training set exclusively sourced from Reddit. While this provides a sufficient foundation for an initial study, further research is needed to explore the generalizability of our approach across different platforms and contexts.
Finally, while we provide an extensive analysis of model selection, configuration design, and training set variations, our exploration is not exhaustive. Future research could expand on these aspects by experimenting with additional modeling choices and training strategies.

\paragraph{Potential Applications}
Despite its limitations, the classification method we propose holds promise for various applications that merit future research.
First, it can enhance community targeting for activism campaigns by identifying broader audiences receptive to collective action causes on large-scale social media platforms, beyond traditional niche ``activist'' communities. This is particularly relevant on Reddit, where political discussions often occur in non-political subreddits~\cite{rajadesingan2021political}.
Second, our classifiers can generate a large-scale and granular ground truth, which is invaluable for training and testing opinion dynamics models~\cite{lenti2024likelihood}.
These models, critical for understanding collective action dynamics, are often limited by insufficient ground truth data on opinion shifts and the social interactions driving them. Our classifiers address this by identifying relevant interactions indicative of a certain inclination towards collective action and their timing.
Finally, applying our tool to online conversations relevant to social movements and activism organizations could enhance our understanding of the interplay between online activity and offline engagement among crowds~\cite{smith2023digital}.

\section*{Acknowledgments}
We acknowledge the support from the Carlsberg Foundation through the COCOONS project (CF21-0432). The funder had no role in study design, data collection and analysis, decision to publish, or preparation of the manuscript.

\bibliography{aaai25}

\begin{thebibliography}{72}
\providecommand{\natexlab}[1]{#1}

\bibitem[{Aceves and Evans(2024)}]{aceves2024mobilizing}
Aceves, P.; and Evans, J.~A. 2024.
\newblock Mobilizing conceptual spaces: How word embedding models can inform measurement and theory within organization science.
\newblock \emph{Organization Science}, 35(3): 788--814.

\bibitem[{Barthel et~al.(2016)Barthel, Stocking, Holcomb, and Mitchell}]{barthel2016reddit}
Barthel, M.; Stocking, G.; Holcomb, J.; and Mitchell, A. 2016.
\newblock Reddit news users more likely to be male, young and digital in their news preferences.
\newblock \emph{Pew Research Center}, 25.

\bibitem[{Benford and Snow(2000)}]{benford2000framing}
Benford, R.~D.; and Snow, D.~A. 2000.
\newblock Framing processes and social movements: An overview and assessment.
\newblock \emph{Annual review of sociology}, 26(1): 611--639.

\bibitem[{Boulianne and Theocharis(2020)}]{boulianne2020young}
Boulianne, S.; and Theocharis, Y. 2020.
\newblock Young people, digital media, and engagement: A meta-analysis of research.
\newblock \emph{Social science computer review}, 38(2): 111--127.

\bibitem[{Brown, Lowery, and Smith(2022)}]{brown2022opposing}
Brown, O.; Lowery, C.; and Smith, L.~G. 2022.
\newblock How opposing ideological groups use online interactions to justify and mobilise collective action.
\newblock \emph{European Journal of Social Psychology}, 52(7): 1082--1110.

\bibitem[{Bucher and Martini(2024)}]{bucher2024fine}
Bucher, M. J.~J.; and Martini, M. 2024.
\newblock Fine-Tuned'Small'LLMs (Still) Significantly Outperform Zero-Shot Generative AI Models in Text Classification.
\newblock \emph{arXiv preprint arXiv:2406.08660}.

\bibitem[{Cerqueira et~al.(2023)Cerqueira, Torgo, Branco, and Bellinger}]{cerqueira2023automated}
Cerqueira, V.; Torgo, L.; Branco, P.; and Bellinger, C. 2023.
\newblock Automated imbalanced classification via layered learning.
\newblock \emph{Machine Learning}, 112(6): 2083--2104.

\bibitem[{Choi et~al.(2020)Choi, Aiello, Varga, and Quercia}]{choi2020ten}
Choi, M.; Aiello, L.~M.; Varga, K.~Z.; and Quercia, D. 2020.
\newblock Ten social dimensions of conversations and relationships.
\newblock In \emph{Proceedings of The Web Conference 2020}, 1514--1525.

\bibitem[{Choi et~al.(2023)Choi, Pei, Kumar, Shu, and Jurgens}]{choi2023llms}
Choi, M.; Pei, J.; Kumar, S.; Shu, C.; and Jurgens, D. 2023.
\newblock Do llms understand social knowledge? evaluating the sociability of large language models with socket benchmark.
\newblock \emph{arXiv preprint arXiv:2305.14938}.

\bibitem[{Coff{\'e} and Bolzendahl(2010)}]{coffe2010same}
Coff{\'e}, H.; and Bolzendahl, C. 2010.
\newblock Same game, different rules? Gender differences in political participation.
\newblock \emph{Sex roles}, 62: 318--333.

\bibitem[{Dettmers et~al.(2024)Dettmers, Pagnoni, Holtzman, and Zettlemoyer}]{dettmers2024qlora}
Dettmers, T.; Pagnoni, A.; Holtzman, A.; and Zettlemoyer, L. 2024.
\newblock Qlora: Efficient finetuning of quantized llms.
\newblock \emph{Advances in Neural Information Processing Systems}, 36.

\bibitem[{Dubey et~al.(2024)Dubey, Jauhri, Pandey, Kadian, Al-Dahle, Letman, Mathur, Schelten, Yang, Fan et~al.}]{dubey2024LLaMa}
Dubey, A.; Jauhri, A.; Pandey, A.; Kadian, A.; Al-Dahle, A.; Letman, A.; Mathur, A.; Schelten, A.; Yang, A.; Fan, A.; et~al. 2024.
\newblock The llama 3 herd of models.
\newblock \emph{arXiv preprint arXiv:2407.21783}.

\bibitem[{Earl, Maher, and Elliott(2017)}]{earl2017youth}
Earl, J.; Maher, T.~V.; and Elliott, T. 2017.
\newblock Youth, activism, and social movements.
\newblock \emph{Sociology Compass}, 11(4): e12465.

\bibitem[{Ebrahimi et~al.(2017)Ebrahimi, Rao, Lowd, and Dou}]{ebrahimi2017hotflip}
Ebrahimi, J.; Rao, A.; Lowd, D.; and Dou, D. 2017.
\newblock Hotflip: White-box adversarial examples for text classification.
\newblock \emph{arXiv preprint arXiv:1712.06751}.

\bibitem[{Erseghe et~al.(2023)Erseghe, Badia, D{\v{z}}anko, Formanowicz, Nikadon, and Suitner}]{erseghe2023projection}
Erseghe, T.; Badia, L.; D{\v{z}}anko, L.; Formanowicz, M.; Nikadon, J.; and Suitner, C. 2023.
\newblock Projection of Socio-Linguistic markers in a semantic context and its application to online social networks.
\newblock \emph{Online Social Networks and Media}, 37: 100271.

\bibitem[{FORCE11(2020)}]{fair}
FORCE11. 2020.
\newblock The FAIR Data principles.
\newblock \url{https://force11.org/info/the-fair-data-principles/}.
\newblock Accessed: 2024-01-05.

\bibitem[{Furlong and Vignoles(2021)}]{furlong2021social}
Furlong, C.; and Vignoles, V.~L. 2021.
\newblock Social identification in collective climate activism: Predicting participation in the environmental movement, extinction rebellion.
\newblock \emph{Identity}, 21(1): 20--35.

\bibitem[{Gamson(1975)}]{gamson1975strategy}
Gamson, W.~A. 1975.
\newblock The Strategy of Social Protest.

\bibitem[{Gebru et~al.(2021)Gebru, Morgenstern, Vecchione, Vaughan, Wallach, Iii, and Crawford}]{gebru2021datasheets}
Gebru, T.; Morgenstern, J.; Vecchione, B.; Vaughan, J.~W.; Wallach, H.; Iii, H.~D.; and Crawford, K. 2021.
\newblock Datasheets for datasets.
\newblock \emph{Communications of the ACM}, 64(12): 86--92.

\bibitem[{Gerbaudo(2012)}]{gerbaudo2012tweets}
Gerbaudo, P. 2012.
\newblock \emph{Tweets and the streets: Social media and contemporary activism}.
\newblock Pluto Press.

\bibitem[{Ghaddar et~al.(2021)Ghaddar, Langlais, Rashid, and Rezagholizadeh}]{ghaddar2021context}
Ghaddar, A.; Langlais, P.; Rashid, A.; and Rezagholizadeh, M. 2021.
\newblock Context-aware adversarial training for name regularity bias in named entity recognition.
\newblock \emph{Transactions of the Association for Computational Linguistics}, 9: 586--604.

\bibitem[{Grootendorst(2022)}]{grootendorst2022bertopic}
Grootendorst, M. 2022.
\newblock BERTopic: Neural topic modeling with a class-based TF-IDF procedure.
\newblock \emph{arXiv preprint arXiv:2203.05794}.

\bibitem[{Gulliver, Fielding, and Louis(2021)}]{gulliver2021assessing}
Gulliver, R.; Fielding, K.~S.; and Louis, W.~R. 2021.
\newblock Assessing the mobilization potential of environmental advocacy communication.
\newblock \emph{Journal of Environmental Psychology}, 74: 101563.

\bibitem[{Hofmann, Sch{\"u}tze, and Pierrehumbert(2022)}]{hofmann2022reddit}
Hofmann, V.; Sch{\"u}tze, H.; and Pierrehumbert, J.~B. 2022.
\newblock The reddit politosphere: a large-scale text and network resource of online political discourse.
\newblock In \emph{Proceedings of the International AAAI Conference on Web and Social Media}, volume~16, 1259--1267.

\bibitem[{Hoyt et~al.(2018)Hoyt, Moss, Burnette, Schieffelin, and Goethals}]{hoyt2018wealth}
Hoyt, C.~L.; Moss, A.~J.; Burnette, J.~L.; Schieffelin, A.; and Goethals, A. 2018.
\newblock Wealth inequality and activism: Perceiving injustice galvanizes social change but perceptions depend on political ideologies.
\newblock \emph{European Journal of Social Psychology}, 48(1): O81--O90.

\bibitem[{Huang et~al.(2025)Huang, Yi, Yu, and Xu}]{huang2025unmasking}
Huang, T.; Yi, J.; Yu, P.; and Xu, X. 2025.
\newblock Unmasking Digital Falsehoods: A Comparative Analysis of LLM-Based Misinformation Detection Strategies.
\newblock \emph{arXiv preprint arXiv:2503.00724}.

\bibitem[{Israeli, Kremiansky, and Tsur(2022)}]{israeli2022must}
Israeli, A.; Kremiansky, A.; and Tsur, O. 2022.
\newblock This must be the place: Predicting engagement of online communities in a large-scale distributed campaign.
\newblock In \emph{Proceedings of the ACM Web Conference 2022}, 1673--1684.

\bibitem[{Kim et~al.(2023)Kim, Cha, Kim, and Park}]{kim2023understanding}
Kim, S.; Cha, J.; Kim, D.; and Park, E. 2023.
\newblock Understanding mental health issues in different subdomains of social networking services: computational analysis of text-based Reddit posts.
\newblock \emph{Journal of Medical Internet Research}, 25: e49074.

\bibitem[{Klandermans(1984)}]{klandermans1984mobilization}
Klandermans, B. 1984.
\newblock Mobilization and participation: Social-psychological expansisons of resource mobilization theory.
\newblock \emph{American sociological review}, 583--600.

\bibitem[{Klandermans(1988)}]{klandermans1988formation}
Klandermans, B. 1988.
\newblock The formation and mobilization of consensus.
\newblock \emph{International social movement research}, 1(1): 173--196.

\bibitem[{Klandermans and Oegema(1987)}]{klandermans1987potentials}
Klandermans, B.; and Oegema, D. 1987.
\newblock Potentials, networks, motivations, and barriers: Steps towards participation in social movements.
\newblock \emph{American sociological review}, 519--531.

\bibitem[{Klandermans and Stekelenburg(2014)}]{klandermans2014people}
Klandermans, B.; and Stekelenburg, J.~V. 2014.
\newblock Why people don't participate in collective action.
\newblock \emph{Journal of Civil Society}, 10(4): 341--352.

\bibitem[{Kumar et~al.(2018)Kumar, Hamilton, Leskovec, and Jurafsky}]{kumar2018community}
Kumar, S.; Hamilton, W.~L.; Leskovec, J.; and Jurafsky, D. 2018.
\newblock Community interaction and conflict on the web.
\newblock In \emph{Proceedings of the 2018 world wide web conference}, 933--943.

\bibitem[{Kumarage, Bhattacharjee, and Garland(2024)}]{kumarage2024harnessing}
Kumarage, T.; Bhattacharjee, A.; and Garland, J. 2024.
\newblock Harnessing artificial intelligence to combat online hate: Exploring the challenges and opportunities of large language models in hate speech detection.
\newblock \emph{arXiv preprint arXiv:2403.08035}.

\bibitem[{Lan et~al.(2024)Lan, Gao, Jin, and Li}]{lan2024stance}
Lan, X.; Gao, C.; Jin, D.; and Li, Y. 2024.
\newblock Stance detection with collaborative role-infused llm-based agents.
\newblock In \emph{Proceedings of the international AAAI conference on web and social media}, volume~18, 891--903.

\bibitem[{Lenti et~al.(2025)Lenti, Aiello, Monti, and Morales}]{lenti2025causal}
Lenti, J.; Aiello, L.~M.; Monti, C.; and Morales, G. D.~F. 2025.
\newblock Causal Modeling of Climate Activism on Reddit.
\newblock In \emph{Proceedings of the ACM on Web Conference 2025}, 590--600.

\bibitem[{Lenti, Monti, and De~Francisci~Morales(2024)}]{lenti2024likelihood}
Lenti, J.; Monti, C.; and De~Francisci~Morales, G. 2024.
\newblock Likelihood-based methods improve parameter estimation in opinion dynamics models.
\newblock In \emph{Proceedings of the 17th ACM International Conference on Web Search and Data Mining}, 350--359.

\bibitem[{Liu et~al.(2021)Liu, Liu, Lu, Welleck, West, Bras, Choi, and Hajishirzi}]{liu2021generated}
Liu, J.; Liu, A.; Lu, X.; Welleck, S.; West, P.; Bras, R.~L.; Choi, Y.; and Hajishirzi, H. 2021.
\newblock Generated knowledge prompting for commonsense reasoning.
\newblock \emph{arXiv preprint arXiv:2110.08387}.

\bibitem[{Liu(2019)}]{liu2019roberta}
Liu, Y. 2019.
\newblock Roberta: A robustly optimized bert pretraining approach.
\newblock \emph{arXiv preprint arXiv:1907.11692}, 364.

\bibitem[{Liu et~al.(2023)Liu, Yao, Ton, Zhang, Cheng, Klochkov, Taufiq, and Li}]{liu2023trustworthy}
Liu, Y.; Yao, Y.; Ton, J.-F.; Zhang, X.; Cheng, R. G.~H.; Klochkov, Y.; Taufiq, M.~F.; and Li, H. 2023.
\newblock Trustworthy LLMs: A survey and guideline for evaluating large language models' alignment.
\newblock \emph{arXiv preprint arXiv:2308.05374}.

\bibitem[{Lucchini et~al.(2022)Lucchini, Aiello, Alessandretti, De~Francisci~Morales, Starnini, and Baronchelli}]{lucchini2022reddit}
Lucchini, L.; Aiello, L.~M.; Alessandretti, L.; De~Francisci~Morales, G.; Starnini, M.; and Baronchelli, A. 2022.
\newblock From Reddit to Wall Street: The role of committed minorities in financial collective action.
\newblock \emph{Royal Society Open Science}, 9(4): 211488.

\bibitem[{Luo, Card, and Jurafsky(2020)}]{luo2020detecting}
Luo, Y.; Card, D.; and Jurafsky, D. 2020.
\newblock Detecting stance in media on global warming.
\newblock \emph{arXiv preprint arXiv:2010.15149}.

\bibitem[{M{\o}ller and Aiello(2024)}]{moller2024prompt}
M{\o}ller, A.~G.; and Aiello, L.~M. 2024.
\newblock Prompt Refinement or Fine-tuning? Best Practices for using LLMs in Computational Social Science Tasks.
\newblock \emph{arXiv preprint arXiv:2408.01346}.

\bibitem[{MySociety(2024)}]{ukdebates}
MySociety. 2024.
\newblock UK Parlament Debates.
\newblock \url{https://www.theyworkforyou.com/pwdata/scrapedxml/}.
\newblock Accessed: 2024-12-15.

\bibitem[{Møller et~al.(2024)Møller, Pera, Dalsgaard, and Aiello}]{moller24parrot}
Møller, A.~G.; Pera, A.; Dalsgaard, J.~A.; and Aiello, L.~M. 2024.
\newblock The Parrot Dilemma: Human-Labeled vs. LLM-augmented Data in Classification Tasks.
\newblock In \emph{Proceedings of the European of the Association of Computational Linguistics}, EACL'24. ACL.

\bibitem[{Ostrom(2010)}]{ostrom2010analyzing}
Ostrom, E. 2010.
\newblock Analyzing collective action.
\newblock \emph{Agricultural economics}, 41: 155--166.

\bibitem[{Oswald and Bright(2022)}]{oswald2022climate}
Oswald, L.; and Bright, J. 2022.
\newblock How do climate change skeptics engage with opposing views online? Evidence from a major climate change skeptic forum on Reddit.
\newblock \emph{Environmental Communication}, 16(6): 805--821.

\bibitem[{Otto et~al.(2020)Otto, Donges, Cremades, Bhowmik, Hewitt, Lucht, Rockstr{\"o}m, Allerberger, McCaffrey, Doe et~al.}]{otto2020social}
Otto, I.~M.; Donges, J.~F.; Cremades, R.; Bhowmik, A.; Hewitt, R.~J.; Lucht, W.; Rockstr{\"o}m, J.; Allerberger, F.; McCaffrey, M.; Doe, S.~S.; et~al. 2020.
\newblock Social tipping dynamics for stabilizing Earth’s climate by 2050.
\newblock \emph{Proceedings of the National Academy of Sciences}, 117(5): 2354--2365.

\bibitem[{Park, Hong, and Kim(2020)}]{park2020methodology}
Park, K.; Hong, J.~S.; and Kim, W. 2020.
\newblock A methodology combining cosine similarity with classifier for text classification.
\newblock \emph{Applied Artificial Intelligence}, 34(5): 396--411.

\bibitem[{Parsa et~al.(2022)Parsa, Shi, Xu, Yim, Yin, and Golab}]{parsa2022analyzing}
Parsa, M.~S.; Shi, H.; Xu, Y.; Yim, A.; Yin, Y.; and Golab, L. 2022.
\newblock Analyzing climate change discussions on reddit.
\newblock In \emph{2022 International Conference on Computational Science and Computational Intelligence (CSCI)}, 826--832. IEEE.

\bibitem[{Pera and Aiello(2024)}]{pera2024narratives}
Pera, A.; and Aiello, L.~M. 2024.
\newblock Narratives of Collective Action in YouTube’s Discourse on Veganism.
\newblock In \emph{Proceedings of the International AAAI Conference on Web and Social Media}, volume~18, 1220--1236.

\bibitem[{Pera, Morales, and Aiello(2023)}]{pera2023measuring}
Pera, A.; Morales, G. d.~F.; and Aiello, L.~M. 2023.
\newblock Measuring Behavior Change with Observational Studies: a Review.
\newblock \emph{arXiv preprint arXiv:2310.19951}.

\bibitem[{Rafailov et~al.(2024)Rafailov, Sharma, Mitchell, Manning, Ermon, and Finn}]{rafailov2024direct}
Rafailov, R.; Sharma, A.; Mitchell, E.; Manning, C.~D.; Ermon, S.; and Finn, C. 2024.
\newblock Direct preference optimization: Your language model is secretly a reward model.
\newblock \emph{Advances in Neural Information Processing Systems}, 36.

\bibitem[{Rajadesingan, Budak, and Resnick(2021)}]{rajadesingan2021political}
Rajadesingan, A.; Budak, C.; and Resnick, P. 2021.
\newblock Political discussion is abundant in non-political subreddits (and less toxic).
\newblock In \emph{Proceedings of the International AAAI Conference on Web and Social Media}, volume~15, 525--536.

\bibitem[{Reimers(2019)}]{reimers2019sentence}
Reimers, N. 2019.
\newblock Sentence-BERT: Sentence Embeddings using Siamese BERT-Networks.
\newblock \emph{arXiv preprint arXiv:1908.10084}.

\bibitem[{Rogers, Kovaleva, and Rumshisky(2019)}]{rogers2019calls}
Rogers, A.; Kovaleva, O.; and Rumshisky, A. 2019.
\newblock Calls to action on social media: Detection, social impact, and censorship potential.
\newblock In \emph{Proceedings of the Second Workshop on Natural Language Processing for Internet Freedom: Censorship, Disinformation, and Propaganda}, 36--44.

\bibitem[{Smith, McGarty, and Thomas(2018)}]{smith2018after}
Smith, L.~G.; McGarty, C.; and Thomas, E.~F. 2018.
\newblock After Aylan Kurdi: How tweeting about death, threat, and harm predict increased expressions of solidarity with refugees over time.
\newblock \emph{Psychological science}, 29(4): 623--634.

\bibitem[{Smith et~al.(2023)Smith, Piwek, Hinds, Brown, Chen, and Joinson}]{smith2023digital}
Smith, L.~G.; Piwek, L.; Hinds, J.; Brown, O.; Chen, C.; and Joinson, A. 2023.
\newblock Digital traces of offline mobilization.
\newblock \emph{Journal of Personality and Social Psychology}, 125(3): 496.

\bibitem[{Snow et~al.(1986)Snow, Rochford~Jr, Worden, and Benford}]{snow1986frame}
Snow, D.~A.; Rochford~Jr, E.~B.; Worden, S.~K.; and Benford, R.~D. 1986.
\newblock Frame alignment processes, micromobilization, and movement participation.
\newblock \emph{American sociological review}, 464--481.

\bibitem[{Thomas et~al.(2022)Thomas, Duncan, McGarty, Louis, and Smith}]{thomas2022mobilise}
Thomas, E.~F.; Duncan, L.; McGarty, C.; Louis, W.~R.; and Smith, L.~G. 2022.
\newblock MOBILISE: A higher-order integration of collective action research to address global challenges.
\newblock \emph{Political Psychology}, 43: 107--164.

\bibitem[{Treen et~al.(2022)Treen, Williams, O’Neill, and Coan}]{treen2022discussion}
Treen, K.; Williams, H.; O’Neill, S.; and Coan, T.~G. 2022.
\newblock Discussion of climate change on Reddit: Polarized discourse or deliberative debate?
\newblock \emph{Environmental Communication}, 16(5): 680--698.

\bibitem[{Van~Stekelenburg and Klandermans(2013)}]{van2013social}
Van~Stekelenburg, J.; and Klandermans, B. 2013.
\newblock The social psychology of protest.
\newblock \emph{Current Sociology}, 61(5-6): 886--905.

\bibitem[{Van~Vugt(2009)}]{van2009averting}
Van~Vugt, M. 2009.
\newblock Averting the tragedy of the commons: Using social psychological science to protect the environment.
\newblock \emph{Current Directions in Psychological Science}, 18(3): 169--173.

\bibitem[{Van~Zomeren(2013)}]{van2013four}
Van~Zomeren, M. 2013.
\newblock Four core social-psychological motivations to undertake collective action.
\newblock \emph{Social and Personality Psychology Compass}, 7(6): 378--388.

\bibitem[{Van~Zomeren and Iyer(2009)}]{van2009introduction}
Van~Zomeren, M.; and Iyer, A. 2009.
\newblock Introduction to the social and psychological dynamics of collective action.

\bibitem[{Van~Zomeren, Postmes, and Spears(2008)}]{van2008toward}
Van~Zomeren, M.; Postmes, T.; and Spears, R. 2008.
\newblock Toward an integrative social identity model of collective action: a quantitative research synthesis of three socio-psychological perspectives.
\newblock \emph{Psychological bulletin}, 134(4): 504.

\bibitem[{Waller and Anderson(2019)}]{waller2019generalists}
Waller, I.; and Anderson, A. 2019.
\newblock Generalists and specialists: Using community embeddings to quantify activity diversity in online platforms.
\newblock In \emph{The World Wide Web Conference}, 1954--1964.

\bibitem[{Waller and Anderson(2021)}]{waller2021quantifying}
Waller, I.; and Anderson, A. 2021.
\newblock Quantifying social organization and political polarization in online platforms.
\newblock \emph{Nature}, 600(7888): 264--268.

\bibitem[{Wright(2009)}]{wright2009next}
Wright, S.~C. 2009.
\newblock The next generation of collective action research.
\newblock \emph{Journal of social Issues}, 65(4): 859--879.

\bibitem[{Wright, Taylor, and Moghaddam(1990)}]{wright1990responding}
Wright, S.~C.; Taylor, D.~M.; and Moghaddam, F.~M. 1990.
\newblock Responding to membership in a disadvantaged group: From acceptance to collective protest.
\newblock \emph{Journal of personality and social psychology}, 58(6): 994.

\bibitem[{Yasseri et~al.(2016)Yasseri, Margetts, John, and Hale}]{yasseri2016political}
Yasseri, T.; Margetts, H.; John, P.; and Hale, S. 2016.
\newblock \emph{Political turbulence: How social media shape collective action}.
\newblock Princeton University Press.

\bibitem[{Ziems et~al.(2024)Ziems, Held, Shaikh, Chen, Zhang, and Yang}]{ziems2024can}
Ziems, C.; Held, W.; Shaikh, O.; Chen, J.; Zhang, Z.; and Yang, D. 2024.
\newblock Can large language models transform computational social science?
\newblock \emph{Computational Linguistics}, 50(1): 237--291.

\end{thebibliography}

\clearpage

\section{Paper Checklist}

\begin{enumerate}

\item For most authors...
\begin{enumerate}
    \item  Would answering this research question advance science without violating social contracts, such as violating privacy norms, perpetuating unfair profiling, exacerbating the socio-economic divide, or implying disrespect to societies or cultures?
    \answerYes{Yes.}
  \item Do your main claims in the abstract and introduction accurately reflect the paper's contributions and scope?
    \answerYes{Yes.}
   \item Do you clarify how the proposed methodological approach is appropriate for the claims made? 
    \answerYes{Yes, in the ``Methodological Framework" section.}
   \item Do you clarify what are possible artifacts in the data used, given population-specific distributions?
    \answerYes{Yes, we discuss potential biases in our data in the ``Discussion and Conclusion'' section.}
  \item Did you describe the limitations of your work?
    \answerYes{Yes, limitations are presented and discussed in the ``Discussion and Conclusion'' section.}
  \item Did you discuss any potential negative societal impacts of your work?
    \answerYes{Yes, we address negative societal impact in the ``Ethical considerations'' section.}
      \item Did you discuss any potential misuse of your work?
    \answerYes{Yes, we discuss potential misuse in the ```Ethics Statement'' section.}
    \item Did you describe steps taken to prevent or mitigate potential negative outcomes of the research, such as data and model documentation, data anonymization, responsible release, access control, and the reproducibility of findings?
    \answerYes{Yes, we share the steps and our framework in the ``Methodological Framework" section. Full experimental details (code) are provided as a shared anonymed link. Moreover, data does not contain user-specific information, only anonymized measures (see DataSheet).}
  \item Have you read the ethics review guidelines and ensured that your paper conforms to them?
    \answerYes{Yes.}
\end{enumerate}

\item Additionally, if your study involves hypotheses testing...
\begin{enumerate}
  \item Did you clearly state the assumptions underlying all theoretical results?
    \answerYes{Yes, we frame our experiments within the scope of previous research in the ``Theoretical Framework" and ``Methodological Framework" sections.}
  \item Have you provided justifications for all theoretical results?
    \answerYes{Yes, justifications are provided in the ``Results" and ``Discussion and Conclusion" sections.}
  \item Did you discuss competing hypotheses or theories that might challenge or complement your theoretical results?
    \answerYes{Yes, other approaches in terms of theoretical frameworks and the motivation behind our choice are described in the ``Methodological Framework" and the ``Related Work" sections.}
  \item Have you considered alternative mechanisms or explanations that might account for the same outcomes observed in your study?
    \answerYes{Yes, in the ``Results" and ``Discussion and Conclusion" sections we explore potential confounders of our results and highlight limitations of our work.}
  \item Did you address potential biases or limitations in your theoretical framework?
    \answerYes{Yes, these are addressed in the ``Discussion and Conclusion" section.}
  \item Have you related your theoretical results to the existing literature in social science?
    \answerYes{Yes, we directly relate our work to previous work in Computational Social Science and Social Psychology theories, as detailed in ``Introduction", ``Theoretical Framework", and ``Related Work".}
  \item Did you discuss the implications of your theoretical results for policy, practice, or further research in the social science domain?
    \answerYes{Yes, these are discussed in the ``Discussion and Conclusion" section.}
\end{enumerate}

\item Additionally, if you are including theoretical proofs...
\begin{enumerate}
  \item Did you state the full set of assumptions of all theoretical results?
    \answerNA{NA.}
	\item Did you include complete proofs of all theoretical results?
    \answerNA{NA.}
\end{enumerate}

\item Additionally, if you ran machine learning experiments...
\begin{enumerate}
  \item Did you include the code, data, and instructions needed to reproduce the main experimental results (either in the supplemental material or as a URL)?
    \answerYes{Yes. Included link to anonymized shared repository (see ``Introduction'').}
  \item Did you specify all the training details (e.g., data splits, hyperparameters, how they were chosen)?
    \answerYes{Yes, whenever necessary. Described in ``Methodological Framework", details are given in the code provided in the anonymized shared repository.}
     \item Did you report error bars (e.g., with respect to the random seed after running experiments multiple times)?
    \answerYes{Yes, whenever relevant. In particular, as boxplots showing the variability of fraction of comments expressing participation in collective action within subreddits--Figure 2.}
    \item Did you include the total amount of compute and the type of resources used (e.g., type of GPUs, internal cluster, or cloud provider)?
    \answerYes{Yes, see ``Results''.}
     \item Do you justify how the proposed evaluation is sufficient and appropriate to the claims made? 
    \answerYes{Yes, specified in the ``Methodological Framework" and ``Results" section.}
     \item Do you discuss what is ``the cost`` of misclassification and fault (in)tolerance?
    \answerYes{Yes, in the ``Results" section}
  
\end{enumerate}

\item Additionally, if you are using existing assets (e.g., code, data, models) or curating/releasing new assets, \textbf{without compromising anonymity}...
\begin{enumerate}
  \item If your work uses existing assets, did you cite the creators?
    \answerYes{Yes, creators of models used are cited in the ``Theoretical Framework", ``Methodological Framework" and ``Results" sections.}
  \item Did you mention the license of the assets?
    \answerYes{Yes, see Appendix A.}
  \item Did you include any new assets in the supplemental material or as a URL?
    \answerYes{Yes, annotated data is included in the anonymized shared repository linked in the ``Introduction''.}
  \item Did you discuss whether and how consent was obtained from people whose data you're using/curating?
    \answerNA{NA.}
  \item Did you discuss whether the data you are using/curating contains personally identifiable information or offensive content?
    \answerYes{Yes. Our data does not include PII. We discuss misinformation and offensive content in the ``Ethics Statement" section.}
\item If you are curating or releasing new datasets, did you discuss how you intend to make your datasets FAIR (see \citet{fair})?
\answerYes{Yes, see DataSheet in the Supplementary Material.}
\item If you are curating or releasing new datasets, did you create a Datasheet for the Dataset (see \citet{gebru2021datasheets})? 
\answerYes{Yes, see the Supplementary Material.}
\end{enumerate}

\item Additionally, if you used crowdsourcing or conducted research with human subjects, \textbf{without compromising anonymity}...
\begin{enumerate}
  \item Did you include the full text of instructions given to participants and screenshots?
    \answerYes{Yes, see Appendix B.}
  \item Did you describe any potential participant risks, with mentions of Institutional Review Board (IRB) approvals?
    \answerYes{Yes. Given that crowdworkers are only annotating data, the only potential risks pertain offensive content or misinformation and is illustrated in the ``Ethics Statement".}
  \item Did you include the estimated hourly wage paid to participants and the total amount spent on participant compensation?
    \answerYes{Yes, see ``Methodological Framework".}
   \item Did you discuss how data is stored, shared, and deidentified?
   \answerYes{Yes, see DataSheet in Supplementary Material.}
\end{enumerate}

\end{enumerate}

\newpage
\section{Ethics Statement}
Our analysis and models are based on publicly available social media data, which is commonly used in similar research contexts. We have taken measures to ensure privacy by removing all usernames and user ids that could link comments to specific users, thereby preventing data misuse and privacy leakage.

Given that Reddit content is predominantly user-generated and subject to varying degrees of moderation, the comments analyzed in this study may include messages that do not necessarily advocate for collective action in ethical or respectful ways. We adopted a broad perspective on public discourse on Reddit, without specifically categorizing content that might contain misinformation or harmful expressions. We emphasize the importance of addressing the ethical implications of engaging with online platforms characterized by mixed levels of regulation.

We also acknowledge the potential for our proposed method to be misused, such as in developing strategies for misinformation or propaganda. While we recognize the risks of unintended applications that could harm society, our primary motivation lies in leveraging this work to address significant societal challenges. We remain committed to minimizing misuse and promoting the responsible application of our approach.

\clearpage

\appendix

\setcounter{figure}{0}
\setcounter{table}{0}
\setcounter{equation}{0}
\renewcommand{\thefigure}{A\arabic{figure}}
\renewcommand{\thetable}{A\arabic{table}}
\renewcommand{\theequation}{A\arabic{equation}}

\setcounter{secnumdepth}{2}

\section{Data and Model Training} \label{app:data_model}

\subsection{Subreddits on Activism} \label{app:subreddits}

\begin{table}[ht!]
\centering
\begin{tabular}{l|c|c}
\textbf{Subreddit} & \textbf{N. subscribers}  & \textbf{Creation date}                                  \\ \hline
\textit{antiwork} & 2,814,235 & 2013-08-14 \\ \hline
\textit{EndFPTP} & 11,481 & 2016-07-29 \\ \hline
\textit{neurodiversity} & 75,768 & 2012-04-27 \\ \hline
\textit{Dyslexia} & 27,494 & 2010-11-13 \\ \hline
\textit{AmIFreeToGo} & 52,296 & 2012-10-25 \\ \hline
\textit{FreeSpeech} & 51,552 & 2009-02-15 \\ \hline
\textit{prochoice} & 44,147 & 2010-01-15 \\ \hline
\textit{jillstein} & 6,654 & 2011-04-20 \\ \hline
\textit{LockdownSkepticism} & 54,693 & 2020-03-25 \\ \hline
\textit{publicdefenders} & 10,606 & 2012-10-30 \\ \hline
\textit{SyrianRebels} & 5,224 & 2016-07-30 \\ \hline
\textit{TwoXChromosomes} & 13,666,599 & 2009-07-16 \\ \hline
\textit{againstmensrights} & 24,949 & 2011-07-20 \\ \hline
\textit{europrivacy} & 20,212 & 2014-03-09 \\ \hline
\textit{ADHDers} & 21,574 & 2020-06-01 \\ \hline
\textit{Vegetarianism} & 45,310 & 2008-05-17 \\ \hline
\textit{ShitLiberalsSay} & 184,152 & 2012-07-24 \\ \hline
\textit{WhiteRights} & - & 2010-03-01 \\ \hline
\textit{runaway} & 13,439 & 2011-11-16 \\ \hline
\textit{MayDayStrike} & 46,993 & 2022-01-05 \\ \hline
\textit{Egalitarianism} & 11,267 & 2011-04-05 \\ \hline
\textit{terfisaslur} & - & 2017-02-02 \\ \hline
\textit{humanrights} & 14,471 & 2009-04-16 \\ \hline
\textit{antiwar} & 13,248 & 2008-08-01\\ \hline
\textit{troubledteens} & 43,742 & 2011-03-26 \\ \hline
\textit{AnimalRights} & 25,706 & 2008-04-15 \\ \hline
\textit{HerpesCureResearch} & 24,248 & 2020-04-10 \\ \hline
\textit{FeMRADebates} & 17,302 & 2013-08-06 \\ \hline
\textit{Intactivism}  & 7,086 & 2009-03-07 \\ \hline
\textit{evolutionReddit} & 15,403 & 2012-01-23 \\ \hline
\textit{IronFrontUSA} & 37,519 & 2018-05-20 \\ \hline
\textit{water} & 43,907 & 2008-07-10 \\ \hline
\textit{Political\_Revolution} & 171,161 & 2016-02-03 \\ \hline
\textit{activism} & - & 2008-03-09 \\ \hline
\textit{VeganActivism} & 21,081 & 2013-01-01 \\ \hline
\textit{climate} & 189,199 & 2008-05-07 \\ \hline
\textit{ndp} & 32,909 & 2008-09-08 \\ \hline
\textit{israelexposed} & 54,630 & 2009-09-30 \\ \hline
\textit{MensRights} & 359,895 & 2008-03-19 \\ \hline
\textit{postnationalist} & 3,644 & 2013-08-27 \\ \hline
\textit{badgovnofreedom} & 4,277 & 2013-09-06 \\ \hline
\textit{EvolveSustain} & - & 2012-06-04 \\ \hline
\end{tabular}
\caption{List of subreddits included in the set of reference for training and testing.}
\label{tab:subreddit_list}
\end{table}

Table \ref{tab:subreddit_list} contains the list of subreddits that are considered as reference for the definition of the training and test sets, together with the number of subscribers as of April 2024 (dataset creation date). 
Subreddits with a missing number of subscribers are either banned or private but their data before their change in status can still be retrieved.

When filtering the set of top 40,000 subreddits by activism-related keywords, we excluded subreddits mainly containing non-English content (e.g. \emph{r/écologie}), generic communities related to countries or continents (e.g. \emph{r/unitedstatesofindia}), communities discussing artists or public figures not directly related to activism (e.g. \emph{r/chancetherapper}), subreddits discussing economic theories not specifically related to activism (e.g. \emph{r/mmt\_economics}), communities discussing games, movies or books out of the context of activism (e.g. \emph{r/501st}), subreddits that went private as a protest to Reddit API changes in 2023 and for which it is not possible to retrieve the topical focus from the name (e.g. \emph{r/charcoal}), debate-focused subreddits (e.g. \emph{r/debateavegan}), subreddits in which ``rights'' are mentioned only in connection to copyright (e.g. \emph{r/freesounds}), rental or unemployment rights (e.g. \emph{r/loveforlandlords}, \emph{r/texasunemployment}) but not specifically related to the context of activism, subreddits in which drugs and their use are discussed and are not strictly related to activism on the topics (e.g. \emph{r/stoners}), communities discussing bots rights (e.g. \emph{r/botsrights}).

\subsection{Crowdsourcing Task} \label{app:crowdsource}

\begin{figure}[t!]
    \centering   
    \includegraphics[width=0.49\textwidth]{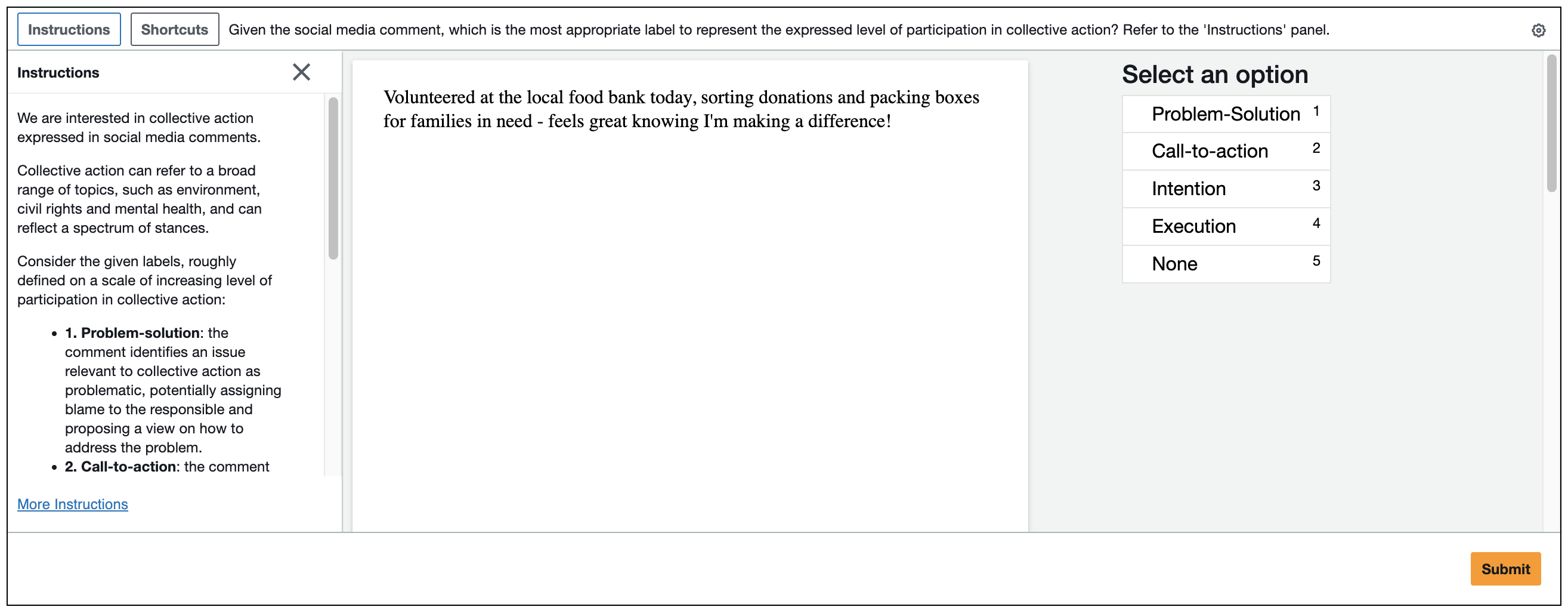}
    \caption{Example screenshot of the annotation task on MTurk.}
    \label{fig:screenshot_crowd}
\end{figure}

We report a screenshot of an example annotation on MTurk in Figure \ref{fig:screenshot_crowd}.

For annotation task instructions, see \texttt{crowdsource\_instructions.pdf} in the GitHub repo \url{https://github.com/ariannap13/extract_collective_action}.

\subsection{Annotation Codebook} \label{app:codebook}
See \texttt{annotation\_codebook.pdf} in the GitHub repo \url{https://github.com/ariannap13/extract_collective_action}.

\renewcommand{\thefigure}{B\arabic{figure}}
\renewcommand{\thetable}{B\arabic{table}}
\renewcommand{\theequation}{B\arabic{equation}}

\begin{table*}[t!]
\centering
\setlength{\tabcolsep}{4pt} 
\begin{tabular}{@{}c|l|cc|cc|cc|cc@{}}
\toprule
\multirow{2}{*}{\textbf{Method}} & \multirow{2}{*}{\textbf{Train. set}} 
& \multicolumn{2}{c|}{\textbf{Participation}} 
& \multicolumn{2}{c|}{\textbf{None}} 
& \multicolumn{2}{c|}{\textbf{Macro Avg.}}  
& \multicolumn{2}{c}{\textbf{Weight. Avg.}}  \\ 

&  & \textbf{P} & \textbf{R} & \textbf{P} & \textbf{R} 
& \textbf{P} & \textbf{R} & \textbf{P} & \textbf{R}  \\ 
\midrule
Dict.  &  & 0.27 & 0.67 & 0.77 & 0.38 & 0.52 & 0.53 & 0.64 & 0.46  \\
Centr. &  & 0.36 & 0.81 & 0.88 & 0.49 & 0.62 & 0.65 & 0.75 & 0.57  \\
\midrule
\multirow{2}{*}{{BERT}}  
& CS   & 0.29 & 0.96 & 0.93 & 0.17 & 0.61 & 0.56 & 0.76 & 0.37  \\
& CS + Syn A & 0.44 & 0.60 & 0.84 & 0.73 & 0.64 & 0.67 & 0.74 & 0.70  \\
\midrule
\multirow{2}{*}{{ZS}}  
& no def & 0.53 & 0.58 & 0.85 & 0.82 & 0.69 & 0.70 & 0.77 & 0.76  \\
& def & 0.53 & 0.66 & 0.87 & 0.80 & 0.70 & 0.73 & 0.78 & 0.76  \\
\midrule
\multirow{4}{*}{{SFT}}  
& CS (no def) & 0.32 & 0.88 & 0.89 & 0.35 & 0.61 & 0.61 & 0.74 & 0.49  \\
& CS (def) & 0.32 & 0.93 & 0.93 & 0.31 & 0.62 & 0.62 & 0.77 & 0.47  \\
& CS + Syn A (no def) & 0.35 & 0.86 & 0.90 & 0.44 & 0.62 & 0.65 & 0.76 & 0.55  \\
& CS + Syn A (def) & 0.36 & 0.90 & 0.93 & 0.43 & 0.64 & 0.67 & 0.78 & 0.55  \\
\midrule
DPO  & CS + Syn A (def) & 0.31 & 0.95 & 0.94 & 0.27 & 0.62 & 0.61 & 0.77 & 0.44  \\
\bottomrule
\end{tabular}
\caption{Precision (\textbf{P}) and recall (\textbf{R}) scores for the binary classification task.}
\label{tab:performance_binary_precision_recall}
\end{table*}

\renewcommand{\thefigure}{A\arabic{figure}}
\renewcommand{\thetable}{A\arabic{table}}
\renewcommand{\theequation}{A\arabic{equation}}

\subsection{Modeling Details} \label{app:model_details}
For all fine-tuned LLM models, we set aside 10\% of the training data as a validation set.

Both RoBERTa classification models, for the binary and the multi-class tasks, are trained for 30 epochs, with a learning rate of $4e^{-5}$ and a train batch size of 16. Given the presence of class imbalance, the model is adjusted with weights computed as $\frac{tot\_sample}{2*[samples\_class_0, ..., samples\_class_n]}$. 
All other parameters are left at their default values and details can be checked in the shared code.

All Llama3 fine-tuned models are used with a temperature set to 0.01. The SFT approach is trained for 20 epochs and the DPO approach for 10 epochs. Both models are set with a learning rate of $2e^{-4}$, a weight decay of 0.001, a \texttt{max\_grad\_norm} of 0.3, and a warmup ratio of 0.03. We set the QLoRa $\alpha$ to 16,
the dropout rate to 0.05, and the matrix rank to 8. The DPO is trained with 2 randomly chosen examples of wrong responses for the multi-class task.  

For the zero-shot LLM, we used enhanced AI prompting. The prompt of reference is reported in Appendix C. The starting point labels definitions are those used in the crowdsourcing annotation campaign and the following is the list of generated labels definitions:
\begin{itemize}
    \item \textbf{Problem-Solution}: The comment highlights an issue and possibly suggests a way to fix it, often naming those responsible.
    \item \textbf{Call-to-Action}: The comment asks readers to take part in a specific activity, effort, or movement.
    \item \textbf{Intention}: The commenter shares their own desire to do something or be involved in solving a particular issue.
    \item \textbf{Execution}: The commenter is describing their personal experience taking direct actions towards a common goal.
    \item \textbf{None}: The comment doesn't fit into one of these categories; its purpose isn't clear or relevant to collective action.
\end{itemize}

\subsection{Licenses}
The \texttt{Llama-3-8B Instruct} model employed for the zero-shot, SFT and DPO approaches is an open-source LLM released under a commercial use license\footnote{\url{https://www.llama.com/llama3/license/}}. 
The \texttt{S-BERT} model for the extraction of sentence embeddings is licensed under the Apache License 2.0., while the \texttt{RoBERTa base} model used for building the BERT-based classifier and the BERTopic model are licensed under a MIT license.
The collective action dictionary by~\citet{smith2018after} is publicly available and so are the stance detection dataset on climate change created by~\cite{luo2020detecting} and the repository of UK parliamentary debates transcripts~\cite{ukdebates}.
The community embeddings and social dimensions similarity data produced by~\citet{waller2021quantifying} are licensed under the CC-BY 4.0 license.

\renewcommand{\thefigure}{B\arabic{figure}}
\renewcommand{\thetable}{B\arabic{table}}
\renewcommand{\theequation}{B\arabic{equation}}

\section{Results} \label{app:results}

\subsection{Additional Performance Metrics}
Table~\ref{tab:performance_binary_precision_recall} reports the metrics of precision and recall for the binary task, while Table~\ref{tab:performance_multiclass_precision_recall} them for the multi-class task.

\begin{table*}[t!]
\centering
\setlength{\tabcolsep}{5pt} 
\begin{tabular}
{@{}c|l|cc|cc|cc|cc|cc|cc|cc}
\toprule
\multirow{3}{*}{\textbf{Method}} & \multirow{3}{*}{\textbf{Train. Set}} 
& \multicolumn{2}{c|}{\textbf{Prob.-Sol.}} 
& \multicolumn{2}{c|}{\textbf{Call-to-A.}} 
& \multicolumn{2}{c|}{\textbf{Intention}} 
& \multicolumn{2}{c|}{\textbf{Execution}} 
& \multicolumn{2}{c|}{\textbf{None}} 
& \multicolumn{2}{c|}{\textbf{\shortstack{Macro \\ Avg.}}} 
& \multicolumn{2}{c}{\textbf{\shortstack{Weigh. \\ Avg.}}}  \\ 
&  & \textbf{P} & \textbf{R} & \textbf{P} & \textbf{R} 
& \textbf{P} & \textbf{R} & \textbf{P} & \textbf{R}  
& \textbf{P} & \textbf{R} & \textbf{P} & \textbf{R} & \textbf{P} & \textbf{R}  \\ 
\midrule
Centr. &  & 0.33 & 0.45 & 0.19 & 0.13 & 0.03 & 0.09 & 0.08 & 0.15 & 0.84 & 0.73 & 0.29 & 0.31 & 0.69 & 0.63 \\
\midrule
\multirow{2}{*}{{BERT}} 
& CS      & 0.30 & 0.47 & 0.52 & 0.62 & 0.13 & 0.09 & 0.67 & 0.15 & 0.84 & 0.73 & 0.49 & 0.41 & 0.72 & 0.66 \\
& CS + Syn A  & 0.30 & 0.43 & 0.63 & 0.62 & 0.12 & 0.27 & 0.27 & 0.31 & 0.84 & 0.73 & 0.43 & 0.47 & 0.71 & 0.66 \\
\midrule
\multirow{2}{*}{{ZS}} 
& no def  & 0.31 & 0.38 & 0.61 & 0.49 & 0.06 & 0.18 & 0.46 & 0.46 & 0.82 & 0.75 & 0.45 & 0.45 & 0.70 & 0.66 \\
& def     & 0.32 & 0.49 & 0.83 & 0.26 & 0.08 & 0.09 & 0.44 & 0.31 & 0.85 & 0.78 & 0.50 & 0.38 & 0.73 & 0.68 \\
\midrule
\multirow{5}{*}{{SFT (def)}} 
& CS        & 0.30 & 0.48 & 0.56 & 0.56 & 0.20 & 0.09 & 0.50 & 0.38 & 0.84 & 0.73 & 0.48 & 0.45 & 0.72 & 0.67 \\
& Ext       & 0.28 & 0.49 & 0.48 & 0.31 & 0.00 & 0.00 & 0.00 & 0.00 & 0.84 & 0.73 & 0.32 & 0.31 & 0.70 & 0.65 \\
& Ext + Syn I/E & 0.29 & 0.47 & 0.44 & 0.36 & 0.22 & 0.18 & 0.43 & 0.23 & 0.84 & 0.73 & 0.44 & 0.39 & 0.71 & 0.65 \\
& CS + Syn I/E & 0.30 & 0.45 & 0.54 & 0.56 & 0.15 & 0.18 & 0.45 & 0.38 & 0.84 & 0.73 & 0.46 & 0.46 & 0.71 & 0.66 \\
& CS + Syn A & 0.31 & 0.46 & 0.64 & 0.64 & 0.24 & 0.36 & 0.55 & 0.46 & 0.84 & 0.73 & 0.51 & 0.53 & 0.72 & 0.67 \\
\midrule
\multirow{5}{*}{{DPO (def)}} 
& CS        & 0.28 & 0.27 & 0.50 & 0.10 & 0.00 & 0.00 & 0.27 & 0.31 & 0.78 & 0.84 & 0.37 & 0.30 & 0.66 & 0.68 \\
& Ext       & 0.28 & 0.46 & 0.47 & 0.41 & 0.00 & 0.00 & 0.00 & 0.00 & 0.84 & 0.74 & 0.32 & 0.32 & 0.70 & 0.65 \\
& Ext + Syn I/E & 0.27 & 0.45 & 0.35 & 0.21 & 0.00 & 0.00 & 0.33 & 0.08 & 0.82 & 0.74 & 0.36 & 0.29 & 0.68 & 0.64 \\
& CS + Syn I/E & 0.31 & 0.46 & 0.72 & 0.54 & 0.18 & 0.27 & 0.44 & 0.62 & 0.84 & 0.74 & 0.50 & 0.52 & 0.72 & 0.67 \\
& CS + Syn A & 0.33 & 0.45 & 0.50 & 0.69 & 0.30 & 0.27 & 0.42 & 0.62 & 0.84 & 0.74 & 0.48 & 0.55 & 0.72 & 0.68 \\
\bottomrule
\end{tabular}
\caption{Precision (\textbf{P}) and recall (\textbf{R}) metrics for the multi-class classification task as a second step of the layered approach.}
\label{tab:performance_multiclass_precision_recall}
\end{table*}

\subsection{Additional Experiments on Test Set}
Table~\ref{tab:performance_multiclass_direct} reports the performance of the classification approaches applied directly to the multi-class task for identifying the nuances of participation in collective action.

\begin{table*}[ht!]
\centering
\begin{tabular}{@{}c|l|ccccccc@{}}
\toprule
\textbf{Method}  &  \textbf{Train. set}     & \textbf{Problem-solution} & \textbf{Call-to-action} & \textbf{Intention} & \textbf{Execution} & \textbf{None} & \textbf{\shortstack{Macro \\ Avg.}} & \textbf{\shortstack{Weight. \\ Avg.}} \\
\midrule
Centr.     &          & 0.37                      & 0.13                   & 0.04               & 0.05                   & 0.38          & 0.2     & 0.36           \\
\midrule
\multirow{2}{*}{{BERT}} 
                          & CS      & 0.36                      & 0.51                    & 0.07               & 0.09                   & 0.41          & 0.29    & 0.40             \\
                          & CS + Syn A       & 0.4                     & 0.49                   & 0.17               & 0.28                  & 0.62       & 0.39  & 0.56            \\
\midrule
\multirow{2}{*}{{ZS}} 
                          & No def                & 0.35                     & 0.51                    & 0.04               & 0.36                   & 0.38          & 0.33      & 0.37          \\
                          & Def                & 0.38             & 0.41                    & 0.12      & 0.27                   & 0.44          & 0.32   & 0.42              \\
\midrule
\multirow{5}{*}{{SFT (def)}} 
                          & CS              & 0.4                      & 0.61                    & 0.08               & 0.33                   & 0.5          & 0.39    & 0.48             \\
                          & Ext.                   & 0.34                      & 0.41                    & 0               & 0.1                   & 0.49          & 0.27       & 0.45         \\
                          & Ext. + Syn I/E & 0.34                      & 0.34                    & 0.11               & 0.33                  & 0.51         & 0.33    & 0.46             \\
                          & CS + Syn I/E & 0.39                      & 0.66                    & 0.2               & 0.47                   & 0.55          & 0.45   & 0.52             \\
                          & CS + Syn A & 0.39                   & 0.56                    & 0.17            & 0.48               & 0.56       & 0.43     & 0.52        \\
\midrule
\multirow{5}{*}{{DPO (def)}} 
                          & CS              & 0.41                      & 0.48                    & 0               & 0.4                   & 0.63          & 0.39   & 0.57               \\
                          & Ext.                   & 0.33                      & 0.22                    & 0.21               & 0                   & 0.48          & 0.25     & 0.43           \\
                          & Ext. + Syn I/E      & 0.35             & 0.28           & 0.29               & 0.32          & 0.66 & 0.38 & 0.58     \\
                          & CS + Syn I/E & 0.37                   & 0.54                    & 0.2            & 0.37                & 0.4      & 0.38 & 0.4            \\
                          & CS + Syn A & 0.39                   & 0                    & 0.23            & 0.42               & 0.51       & 0.31   & 0.46          \\
\bottomrule
\end{tabular}
\caption{F1 scores for the multi-class classification task performed directly.}
\label{tab:performance_multiclass_direct}
\end{table*}

\noindent Table~\ref{tab:performance_multiclass_nodef} reports the performance of the LLM fine-tuning-based classification approaches that do not consider a definition of what a collective action problem is in the prompt. Note that both results of the direct application to the multi-class task and results of the multi-class task as a second step of the layered framework are reported. 

\begin{table*}[ht]
\centering
\begin{tabular}{@{}c|l|ccccccc@{}}
\toprule
\textbf{Method}  & \textbf{Train. set}       & \textbf{Problem-solution} & \textbf{Call-to-action} & \textbf{Intention} & \textbf{Execution} & \textbf{None} & \textbf{\shortstack{Macro \\ Avg.}} & \textbf{\shortstack{Weight. \\ Avg.}} \\
\midrule
\multirow{5}{*}{{\shortstack{SFT\\(direct)}}} 
                          & CS              & 0.42                      & 0.49                    & 0.18               & 0.31                   & 0.6          & 0.4 & 0.55                \\
                          & Ext.                   & 0.35                      & 0.34                    & 0.1               & 0.1                   & 0.5          & 0.28    & 0.45                     \\
                          & Ext. + Syn I/E & 0.35                      & 0.26                    & 0.14               & 0.14                  & 0.53          & 0.29 & 0.48                \\
                          & CS + Syn I/E & 0.4                      & 0.6                    & 0.2               & 0.43                   & 0.62          & 0.45    & 0.57             \\
                          & CS + Syn A & 0.4                   & 0.57                    & 0.15              & 0.45                   & 0.57          & 0.43 & 0.53             \\
\midrule
\multirow{5}{*}{{\shortstack{DPO\\(direct)}}} 
                          & CS              & 0.38                      & 0.38                    & 0.07               & 0.3                   & 0.56          & 0.34 & 0.51                \\
                          & Ext.                   & 0.34                      & 0.32                    & 0.24               & 0.19                   & 0.49          & 0.32        & 0.45                     \\
                          & Ext. + Syn I/E      & 0.34             & 0.3           & 0.22               & 0.22          & 0.58 & 0.33 & 0.51      \\
                          & CS + Syn I/E & 0.4                      & 0.39                   & 0.17           & 0.39                   & 0.59         & 0.39  & 0.54              \\
                          & CS + Syn A & 0.38                      & 0.35                    & 0.1            & 0.47                   & 0.52          & 0.36 & 0.48                \\
\midrule
\multirow{5}{*}{{\shortstack{SFT\\(layer)}}} 
                          & CS              & 0.36                      & 0.45                    & 0               & 0.24                   & 0.78          & 0.37  & 0.67                \\
                          & Ext.                   & 0.36                      & 0.48                    & 0.11               & 0.12                   & 0.78          & 0.37  & 0.67                \\
                          & Ext. + Syn I/E & 0.37                      & 0.45                    & 0.11               & 0.35                  & 0.78          & 0.41 & 0.68                \\
                          & CS + Syn I/E & 0.36                      & 0.56                   & 0.19               & 0.38                   & 0.78          & 0.46 & 0.68                \\
                          & CS + Syn A & 0.36                      & 0.6             & 0.38          & 0.48                   & 0.78          & 0.52 & 0.69                \\
\midrule
\multirow{5}{*}{{\shortstack{DPO\\(layer)}}} 
                          & CS              & 0.38                      & 0.48                    & 0.1               & 0.31                   & 0.78          & 0.41 & 0.41                \\
                          & Ext.                   & 0.34                      & 0.26                    & 0.12               & 0                   & 0.78          & 0.30 & 0.30                \\
                          & Ext. + Syn I/E      & 0.36             & 0.33           & 0.1               & 0.33          & 0.78 & 0.38 & 0.38     \\
                          & CS + Syn I/E & 0.35                   & 0.53                    & 0.13              & 0.45                   & 0.78         & 0.45 & 0.45                \\
                          & CS + Syn A & 0.35                     & 0.64                  & 0.24              & 0.42                   & 0.78          & 0.49 & 0.48                \\
\bottomrule
\end{tabular}
\caption{F1 scores for the multi-class classification task for the LLM fine-tuning approaches. This table considers the multi-class task performed both directly and as a second step of the layered approach, without considering the definition of what a collective action problem is in the prompt.}
\label{tab:performance_multiclass_nodef}
\end{table*}

\subsection{Additional Details on Topic Modeling}

\begin{figure}[t!]
    \centering   
    \includegraphics[width=0.30\textwidth]{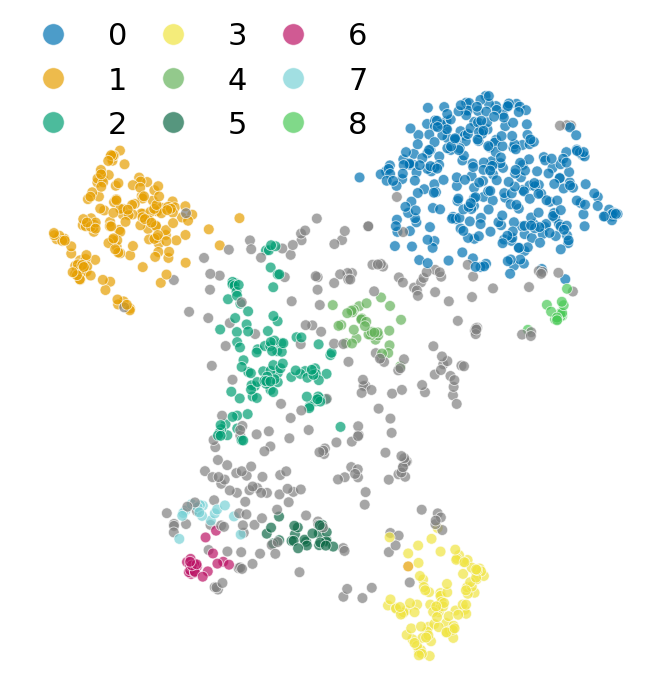}
    \caption{Complete visual representation of the topic modeling application, resulting in 9 topics (plus noise).}
    \label{fig:umap_topics_all}
\end{figure}

Table \ref{tab:topics_full} shows the complete configuration of topics identified in the set involving one thread from \emph{r/intj}, one thread from \emph{r/wallstreetbets}, and one thread from \emph{r/IronFrontUSA}. Since topics 4-8 are quite small in size and can be either considered noise or be attributed to \emph{r/intj} (topic 7, topic embedding similarity with topic 0 about covid and vaccines: 0.527) and \emph{r/IronFrontUSA} (topic 8, topic embedding similarity with topic 1 about voting: 0.628), we focused on topics 0-3 for the sake of simplicity in the visualization reported in the main text. For completeness, we report the same visualization considering all topics in Figure \ref{fig:umap_topics_all}.

\begin{table}[ht!]
\centering
\begin{tabular}{c|l|c}
\textbf{Topic n.} & \textbf{Topic keywords}  & \textbf{N. comments}                                  \\ \hline
\textit{-1} & \shortstack{internet, say,\\thats, yes} & 213 \\ \hline
\textit{0} & \shortstack{covid, effects,\\vaccine, risk} & 332 \\ \hline
\textit{1} & \shortstack{voting, vote,\\democrats, power} & 165 \\ \hline
\textit{2} & \shortstack{science, im,\\read, intj} & 141 \\ \hline
\textit{3} & \shortstack{robinhood, money,\\gme, app}  & 105 \\ \hline
\textit{4} & \shortstack{d*ck, twist,\\it, f*ck} & 31 \\ \hline
\textit{5} & \shortstack{this, fr,\\rarity, baaaa} & 25 \\ \hline
\textit{6} & \shortstack{yes, shit,\\sure, no} & 21 \\ \hline
\textit{7} & \shortstack{booster, shot,\\soon, yes} & 16 \\ \hline
\textit{8} & \shortstack{government, white,\\targeted, person} & 10 \\ \hline
\end{tabular}
\caption{List of topics identified by applying BERTopic on the reference dataset for topic modeling.}
\label{tab:topics_full}
\end{table}

\renewcommand{\thefigure}{C\arabic{figure}}
\renewcommand{\thetable}{C\arabic{table}}
\renewcommand{\theequation}{C\arabic{equation}}

\section{Prompts} \label{app:prompts}

Note that, for the \textbf{Labels Descriptions Generations} prompt, the label definitions to be re-written by the LLM are those presented to crowdworkers (see Section \ref{app:crowdsource}).

\clearpage

\begin{promptbox}{Control Samples Generation/Data Augmentation}
System prompt: You are an advanced AI writer. Your job is to help write examples of social media comments that convey certain dimensions of engagement in collective action, starting from an anchor example.
Prompt: The following anchor example conveys the dimension {label}. {label} is defined by {social\_dimension\_description}. Write 20 new examples that are semantically similar to the anchor example, maintaining the same content structure but slightly varying the topic. Ensure each generated comment retains the core meaning, intent, and key details of the anchor, while introducing minor variations in topic, wording or minor shifts in context. Do not drastically change the main idea or subject matter. Put each generated example on a new line.
Anchor example: {text}
Generated data:
\end{promptbox}

\begin{promptbox}{Labels Definitions Generations} 
System prompt: You are an advanced re-writing AI. You are tasked with re-writing definitions of collective action dimensions to make them easier to understand for a Large Language Model during zero-shot tasks.
Prompt: You have the following knowledge about collective action dimensions that can be expressed in social media comments: {dimdef}. Re-write the definition for each of the dimensions in {labels_dims}. Answer ONLY with 'dimension':'definition'.
Text: {text}
Answer:
\end{promptbox}

\begin{promptbox}{LLM Binary Task (no def)}
Classify whether the social media comment expresses collective action ("1") or not ("0").
A comment is considered to express collective action if fits in any of the following descriptions: 
* The comment highlights an issue and suggests a way to fix it, often naming those responsible.
* The comment asks readers to take part in a specific activity, effort, or movement.
* The commenter shares their own desire to do something or be involved in solving a particular issue.
* The commenter is describing their personal experience taking direct actions towards a common goal.
Return the label "1" or "0" based on the classification.
Comment: {text}
Label: 
\end{promptbox}

\begin{promptbox}{LLM Binary Task (def)}
Classify the following social media comment as either "1" (expressing participation in collective action) or "0" (not expressing participation in collective action).

### Definitions and Criteria:
**Collective Action Problem:** A present issue caused by human actions or decisions that affects a group and can be addressed through individual or collective efforts.

**Participation in collective action**: A comment must clearly reference a collective action problem, social movement, or activism by meeting at least one of the following:
1. The comment identifies the issue as a problem and optionally proposes solutions and/or assigns responsibility.
2. The comment encourages others to take action or join a cause.
3. The comment expresses personal intent to act or current involvement in activism.

### Labeling Instructions:
- Label the comment as "1" if it expresses participation in collective action.
- Label the comment as "0" if it does not express participation in collective action.

### Example of correct output
Comment: "xyz"
Label: 0

Return the label "1" or "0" based on the classification.

Comment: {text}
Label:  
\end{promptbox}

\begin{promptbox}{LLM Multi-class Task (no def)}
You have the following knowledge about collective action dimensions that can be expressed in social media comments: {dim_def}. Classify the following social media comment into one of the dimensions within the list {list(dim_def.keys())}, and return the answer as the corresponding collective action dimension label. 
\end{promptbox}

\begin{promptbox}{LLM Multi-class Task (def)}
You have the following knowledge about levels of participation in collective action that can be expressed in social media comments: {dim_def}. 
            
### Definitions and Criteria:
**Collective Action Problem:** A present issue caused by human actions or decisions that affects a group and can be addressed through individual or collective efforts.

**Participation in collective action**: A comment must clearly reference a collective action problem, social movement, or activism by meeting at least one of the levels in the list {dim_def.keys()}.

Classify the following social media comment into one of the levels within the list {list(dim_def.keys())}. 

### Example of correct output format:
text: xyz
label: None

Return the answer as the corresponding participation in collective action level label.

text: {text}
label: 
\end{promptbox}

\end{document}